\documentclass[journal=mamobx, manuscript=article]{achemso}

\usepackage{eurosym}
\usepackage{enumitem}
\usepackage{tabularx}
\usepackage{graphics}
\usepackage{epsfig}
\usepackage{multirow}
\usepackage{amssymb}
\usepackage{amsmath}
\usepackage{leftidx}
\usepackage{epsfig}
\usepackage{color,soul}
\usepackage{array}
\usepackage{comment}
\usepackage{datetime2}
\usepackage{lscape}
\usepackage{enumitem}

\usepackage[explicit]{titlesec}

\renewcommand{\thesubsection}{\Roman{section}.\arabic{subsection}}

\title{Electrostatic Persistence Length Revisited. \\ I. Theory}
\date{\today}

\author{Artem~M.~Rumyantsev}
\email{rumyantsev@ncsu.edu}
\affiliation{Department of Chemical and Biomolecular Engineering, North Carolina State University, Raleigh, North Carolina 27695-7905, United States}

\author{Alexey~A.~Gavrilov}
\affiliation{Department of Chemical and Biomolecular Engineering, North Carolina State University, Raleigh, North Carolina 27695-7905, United States}

\author{Albert~Johner}
\email{albert.johner@ics-cnrs.unistra.fr}
\affiliation{Institut Charles Sadron, Universit\'{e} de Strasbourg, CNRS UPR22, Strasbourg 67034, France}

\begin{document}

\begin{tocentry}
\includegraphics[height=1.75in]{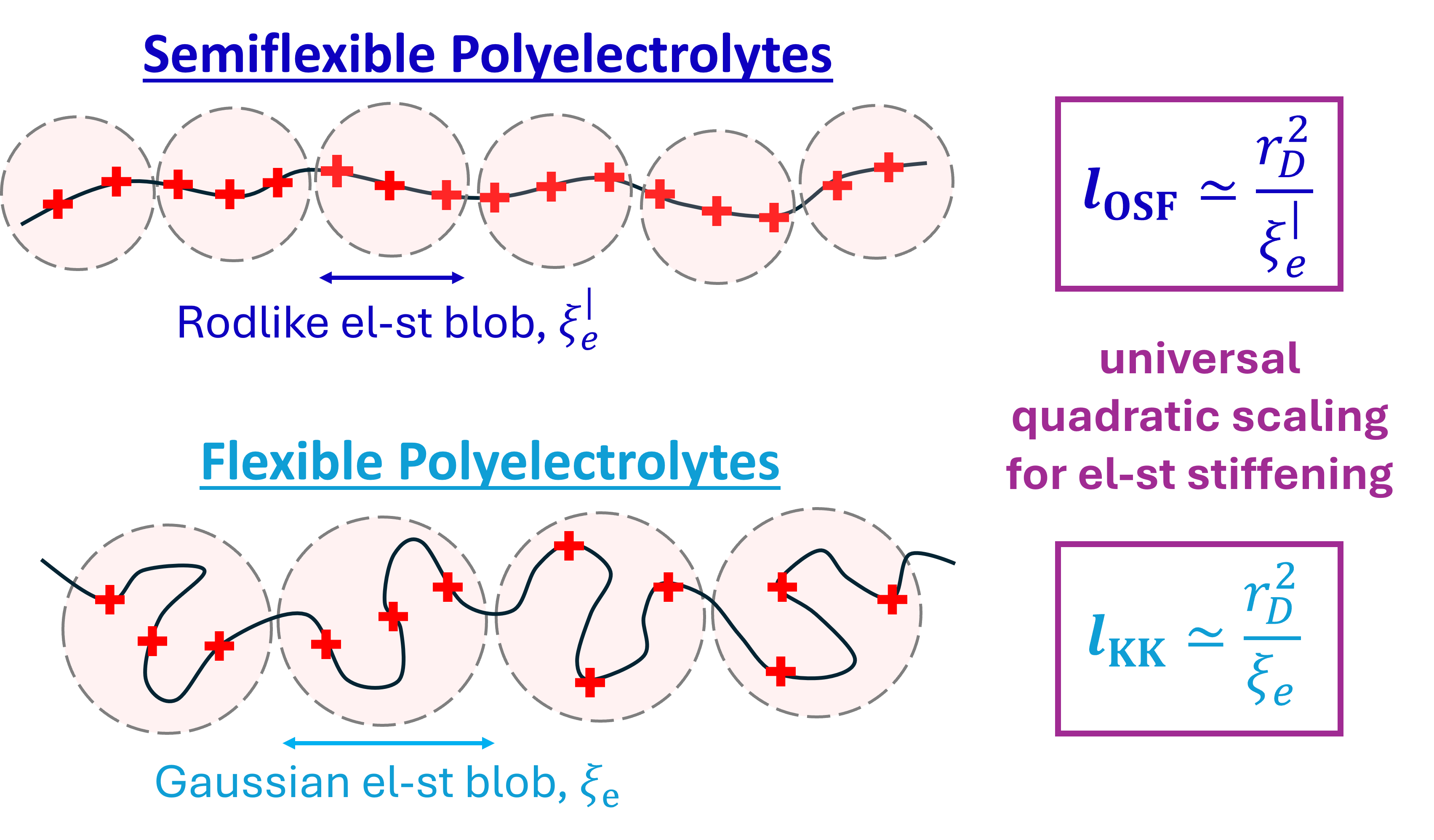}
\end{tocentry}

\begin{abstract}

The problem of single-chain conformations of polyelectrolytes in salt-added dilute solutions, and the associated concept of the electrostatic persistence length $\textbf{l}_\mathrm{e}$, has remained unresolved for decades. To address this challenge, we develop a comprehensive scaling theory and corroborate it with simulations. A unified scaling diagram is constructed that encompasses both flexible and intrinsically semiflexible/stiff polyelectrolytes. Nine distinct scaling regimes are identified, each characterized by different conformational statistics. The concept of the Gaussian electrostatic blob $\xi_\mathrm{e}$ used in the description of flexible polyelectrolytes is extended to the semiflexible case, and the new characteristic length $\xi_\mathrm{e}^{|}$ referred to as the rodlike electrostatic blob is introduced. This enables delineating the important crossovers and showcasing an analogy between semiflexible and flexible chains. The concept of electrostatic excluded volume is also reconsidered and generalized. Upon increasing the salt concentration, i.e., decreasing the Debye length $r_\mathrm{D}$, chain conformations evolve from (i) rodlike stretches to (ii) Gaussian and then (iii) swollen coils with local electrostatic stiffening characterized by $\textbf{l}_\mathrm{e}$, followed by (iv) swollen coils without stiffening but with electrostatic exclude volume, and finally to (v) quasi-neutral Gaussian coils. In regimes of type (ii) and (iii), the electrostatic persistence length scales quadratically with $r_\mathrm{D}$, in agreement with the Odijk–Skolnick–Fixman and Khokhlov–Khachaturian predictions for semiflexible and flexible chains, respectively: $\textbf{l}_\mathrm{\mathrm{OSF}} \simeq r_\mathrm{D}^{2}/\xi_\mathrm{e}^{|}$ and $\textbf{l}_\mathrm{\mathrm{KK}} \simeq r_\mathrm{D}^{2}/\xi_\mathrm{e}$. These asymptotic scalings are confirmed by coarse-grained simulations presented in the accompanying article [II. Simulations and Comparison to Experiment], where the origin of the {\it apparent} linear scaling often observed in experiments is discussed. Our results are vital for achieving a complete and consistent description of dilute and semidilute polyelectrolyte solutions. 

\end{abstract}

\maketitle

\newpage


\section{Introduction}
\label{sec:intro}

The problem of single-chain conformations of polyelectrolytes (PEs) is central to the physics of ion-containing polymers, as it provides the foundation for understanding more complex many-chain systems, such as semidilute solutions. Fifty years ago, in 1976, de Gennes and co-authors pioneered a scaling framework for describing the conformations of single-chain polyelectrolytes in salt-free solutions.~\cite{degennes-1976} However, extending this approach to salt-added solutions has proven to be notoriously difficult. The central challenge arises from the local stiffening of polyelectrolyte chains due to Coulomb interactions between monomers --- an effect commonly quantified by and referred to as the electrostatic persistence length (EPL), further denoted $\textbf{l}_\mathrm{e}$.

Shortly after de Gennes’ seminal work, the problem of electrostatic stiffening of intrinsically \textit{stiff (semiflexible)} single-chain polyelectrolytes in salt-added solutions was considered by Odijk~\cite{odijk-1977} and by Skolnick and Fixman.~\cite{SF-1977} Perturbative calculations in both studies demonstrated that the EPL scales quadratically with the Debye screening length, $\textbf{l}_\mathrm{\mathrm{OSF}} \sim r_\mathrm{D}^{2}$; this result is commonly referred to as the OSF scaling.

Khokhlov and Khachaturian were the first to address the EPL problem in intrinsically \textit{flexible} PEs.~\cite{KK-1982} They combined de Gennes’ scaling picture of a flexible PE as a rodlike string of electrostatic blobs with the OSF calculations performed for such a coarse-grained chain of blobs. It was concluded that the electrostatic persistence length of flexible chains follows a similar quadratic scaling, $\textbf{l}_\mathrm{\mathrm{KK}}  \sim r_\mathrm{D}^{2}$. Although the $r_\mathrm{D}$-independent prefactor in this KK scaling differs, the quadratic dependence on the Debye screening length coincides with that of the OSF result.

However, a decade later, the electrostatic stiffening of flexible PEs was reconsidered by Barrat and Joanny within a variational approach.~\cite{BJ-1993} Their calculations confirmed the quadratic OSF scaling for intrinsically rigid (semiflexible) chains, but predicted a linear dependence of the EPL on the Debye radius for flexible PEs, $\textbf{l}_\mathrm{\mathrm{BJ}} \sim r_\mathrm{D}$. This contradiction between the linear BJ scaling and the quadratic KK scaling has remained a long-standing puzzle in polyelectrolyte theory.

Since then, this issue has been debated in the literature for more than three decades without reaching a broad consensus. Numerous theoretical studies, primarily based on variational calculations, have argued in favor of either quadratic KK scaling~\cite{witten-1995, netz-1999, thirumalai-1999, ullner-2003, netz-andelman-2003, netz-2004, DC-2009, dobrynin-2009} or linear BJ scaling.~\cite{muthu-1987, thirumalai-1995, muthu-1996, muthu-2001, dobrynin-2005} In addition, several works have proposed other alternative laws.~\cite{bratko-1993, liverpool-1997, vilgis-1998, liverpool-2001, podgornik-2001}
Theoretical predictions have been tested by simulations.~\cite{barrat-1993, kremer-1997, everaers-2002, shklovskii-2002, ullner-2002, buehler-2012, prochazka-2012, stevens-2018} However, neither of the answers has yet achieved broad acceptance as a final and comprehensive. 

To a large extent, this lack of consensus stems from methodological issues, particularly concerning the reliability and applicability of the variational techniques employed, which are often essentially uncontrolled. For instance, Manghi and Netz recovered the KK scaling with correct prefactors for flexible PEs, but acknowledged that, for semiflexible chains, their approach yielded the OSF quadratic scaling with an incorrect dependence of the prefactor on the Bjerrum length, for which they could offer no explanation.~\cite{netz-2004} Another example is provided by a series of papers by Dobrynin et al., in which he initially applied the Gibbs–Bogolyubov variational principle to obtain the linear BJ scaling~\cite{dobrynin-2005}, but later recognized that this approach fails to capture long-range orientational correlations and consequently revised his earlier result in favor of the quadratic KK scaling.~\cite{DC-2009, dobrynin-2009} 

Thus, even after almost half a century of research, there is still no broad consensus on the correct scaling of the EPL with the Debye screening length, especially for flexible PE. This state of affairs remains unsatisfactory and deeply challenging, particularly given the apparent simplicity of the minimal problem formulation: What are the conformations of a counterion-free PE whose charges interact via a screened Coulomb (Debye–H\"{u}ckel) potential? 

To achieve a complete understanding of this problem, it is essential to develop a comprehensive scaling picture of PE chains across all relevant regimes, including stiff/flexible chains as well as low/high-salt conditions. Many of the earlier confusions and apparent contradictions stem from the fact that not all relevant characteristic length scales were identified in previous theories, and consequently, not all important crossovers were recognized. As noted above, most earlier studies (except for the KK work) relied primarily on variational calculations rather than on a scaling approach.
Often, both the range of applicability (validity) of such methods~\cite{muthu-1987, muthu-1996, muthu-2001} and their non-trivial technical 
aspects, particularly the integral regularization via the choice of a short-wavelength $q$-cutoff for chain bending fluctuations,~\cite{barrat-1993, thirumalai-1995, witten-1995, thirumalai-1999, netz-2004, dobrynin-2005} can only be understood in a more physically transparent and sounder way within a scaling framework. Moreover, it is extremely difficult, if not impossible, to construct a single-chain functional (Hamiltonian) that is universally applicable across all conditions. This issue becomes particularly significant in light of the main result of the present work: a scaling diagram comprising \text{nine} distinct scaling regimes, each characterized by different conformational statistics of the PE chain.

In this series of papers, we aim to provide a comprehensive resolution of the EPL problem by developing a scaling theory in Part I and corroborating it with coarse-grained simulations in Part II.~\cite{EPL-part2}

The present, theoretical paper is organized as follows. The problem formulation is presented in Section~\ref{sec:setup}. The \textit{local} stiffening of the PE chain, as predicted within the OSF and KK theories, is critically reviewed in Section~\ref{sec:stiffening}. Section~\ref{sec:elst-blob} introduces an important new length scale referred to as the rodlike electrostatic blob. We further discuss the effect of \textit{non-local interactions}. The concept of electrostatic excluded volume, particularly useful at high salt concentrations, is refined in Section~\ref{sec:EEV}. Finally, Section~\ref{sec:diag} presents the main results: a complete diagram of PE conformational regimes together with the corresponding scaling laws for chain structure and dimensions. A concluding discussion is given in Section~\ref{sec:conclusions}. The accompanying paper~\cite{EPL-part2} tests and confirms the key theoretical findings through coarse-grained simulations of the minimal EPL model.

\section{Problem Formulation and Notation}
\label{sec:setup}

The standard problem formulation is as follows. Single polyelectrolyte (PE) chain with the bare Kuhn segment $l_\mathrm{0}$ carries univalent charges, which are spaced equidistantly at the curvilinear distance $A$ from each other. Chain contour length is equal to $L = l_\mathrm{0} N$, and the chain is infinitely thin,~\cite{trizac-2016} that is, all steric effects are completely neglected. In other words, in the absence of charges, the chain would behave as an ideal coil (as in a $\Theta$-solvent). We emphasize that, in our notation, $N$ is not the number of chemical monomers but the number of \textit{bare statistical segments} of the chain.
At the scaling level of accuracy adopted in this work, bare Kuhn segment and bare persistence length are both equal to $l_\mathrm{0}$, so these terms will be used interchangeably. The Bjerrum length of the solution is given by $l_\mathrm{B} = e^2 / \epsilon k_\mathrm{B} T$. The solution contains some salt so the effective interaction potential between charges on the PE has a Debye-H\"{u}ckel form:
\begin{equation}
    \frac{u_\mathrm{12} (r_\mathrm{ij})} {k_\mathrm{B} T} 
    = \frac{e^2}{\epsilon k_\mathrm{B} T} \times \frac{ e^{- r_\mathrm{ij} / r_\mathrm{D}} }{r_\mathrm{ij}} 
    = l_\mathrm{B} \frac{ e^{- r_\mathrm{ij} / r_\mathrm{D}} }{r_\mathrm{ij}} 
\label{eq:1}
\end{equation}
Here $k_\mathrm{B} T$ is the thermal energy, and $r_\mathrm{D}$ is the Debye radius due to small salt ions. Recall that, for the total concentration of salt ions $C_s = C_s^+ + C_s^-$, the Debye radius equals $r_\mathrm{D}^{-2} = 4 \pi l_\mathrm{B} C_s$. It deserves mentioning that eq.~\ref {eq:1} can be derived within the linear response approximation (the random phase approximation, RPA) assuming the salt ions are point-like and interact with the PE and each other only via electrostatic forces.~\cite{BE-1988, RJ-2023} The associated limitations include the requirement of weak charge correlations, $r_\mathrm{D} \gg l_\mathrm{B}$, fulfilled at sufficiently low salt concentrations, $C_s l_\mathrm{B}^3 \ll 1$.~\cite{LL-book} 

The questions to be answered here and in the companion simulations paper~\cite{EPL-part2} relate to the conformational statistics of the PE, namely, to its electrostatic stiffening. How does the presence of charges affect the chain flexibility? What is the apparent persistence length of the PE? Is it possible to consider the PE chain as the neutral chain with an effective total persistence length
\begin{equation}
    \textbf{l} = l_\mathrm{0} + \textbf{l}_\mathrm{e}
\end{equation}
equal to the sum of bare and electrostatic contributions? What are the scaling laws defining $\textbf{l}_\mathrm{e}$ as the function of the 4 other characteristic lengths: $l_\mathrm{B}$, $l_\mathrm{0}$, $A$, and $r_\mathrm{D}$?

It is convenient to express all length scales in the units of the persistence length, $l_\mathrm{0}$. Within the scaling analysis, in which all the numerical coefficients are dropped, this is equivalent to using the units of the statistical segment length equal to $2 l_\mathrm{0}$.~\cite{GK-book} 
Here we implicitly assume that the PE chain is wormlike (persistent) but note that, due to the {\textit scaling level of accuracy} of our analysis, the power-law results are independent of the flexibility mechanism. In other words, the flexibility mechanism is only important locally, but enters global chain properties only through $l_\mathrm{0}$. This also clarifies the physical meaning of $N$ being the number of statistical (Kuhn) segments rather than chemical monomers in the chain. (In principle, one can introduce the chemical monomer size, but the final results are independent of it.~\cite{GK-book}) The dimensionless Bjerrum length expressed in the units of $l_\mathrm{0}$ equals 
\begin{equation}
    u = \frac{l_\mathrm{B}} {l_0} = \frac{e^2} {\epsilon l_\mathrm{0} k_\mathrm{B} T}
\end{equation}
and the fraction/number of ionic monomers in the PE per one bare persistence length is
\begin{equation}
    f = \frac{l_0}{A}
\end{equation}
Note that, up to the numerical coefficient of $2$, this definition coincides with the fraction of ionic statistical segments in flexible PEs widely 
used in the earlier literature.~\cite{BJ-review-1996, DR-2005} At $A \gg l_\mathrm{0}$, their fraction is low, $f \ll 1$. In this limit, the PE chains are referred to as flexible. The opposite case of $A \ll l_\mathrm{0}$ corresponds to PEs with the statistical segment comprising several charges, $f \gg 1$, and these chains are usually semiflexible. Finally, it is convenient to introduce the dimensionless salt concentration as the number of salt ions in the $l_\mathrm{0}^3$ volume of the solution, $c_\mathrm{s} = C_\mathrm{s} l_\mathrm{0}^3$. The Debue radius can then be expressed as
\begin{equation}
r_\mathrm{D} = \left( 4 \pi l_\mathrm{B} C_s \right)^{-1/2} = l_\mathrm{0} \left( 4 \pi u c_s \right)^{-1/2}    
\end{equation}
and the effect of salt can be quantified by the dimensionless $r_\mathrm{D} / l_\mathrm{0}$ ratio.

In what follows, chains (scaling regimes) are classified as \textit{flexible} and \textit{semiflexible} based on neither the bare value of the persistence length $l_\mathrm{0}$ nor the ratio between the Kuhn segment and the monomer diameter, $l_\mathrm{0} / d$, because the latter is assumed to be zero, $ d \to 0$. Instead, we distinguish between them based on their \textit{salt-free local conformational statistics} at lengths just above the Kuhn segment: Gaussian for flexible PEs and rodlike for semiflexible PEs, as shown in Figure~\ref{fig:A-blobs}. Indeed, in the absence of salt, flexible chains remain Gaussian at intermediate lengths above $l_\mathrm{0}$ but not exceeding the Gaussian (i.e., $\Theta$-solvent) electrostatic blob size~\cite{degennes-1976, BJ-review-1996, DR-2005, D-2020}
\begin{equation}
    \xi_\mathrm{e} \simeq l_\mathrm{0} \left( u f^2 \right)^{-1/3} 
\label{elst-blob}
\end{equation}
and exhibit rodlike statistics at larger scales. This representation remains valid as long as $\xi_\mathrm{e} \gg l_\mathrm{0}$ (see Figure~\ref{fig:A-blobs}b). The window of the intermediate Gaussian statistics at $l_\mathrm{0} < r < \xi_\mathrm{e}$ lengths disappears when $\xi_\mathrm{e} \simeq l_\mathrm{0}$, which represents the crossover between semiflexible and flexible regimes. This crossover can also be written as 
\begin{equation}
    u f^2 = \frac{l_\mathrm{B} l_0}{A^2} \simeq  1
\label{cross:stiff}
\end{equation}
and is shown in violet in the scaling diagram of Figure~\ref{fig:1}. When the value of this parameter is much higher than unity, PE demonstrates rodlike statistics at all length scales, and these chains are termed semiflexible (see Figure~\ref{fig:A-blobs}a). Therefore, the dimensionless number $u f^2$ is the first key parameter of the problem. Eq. 7 implies that flexible and semiflexible chains may alternatively be referred to as weakly and strongly charged. To avoid confusion, we use only the former terminology --- flexible and semiflexible PEs --- throughout the paper.

Note that, in the absence of Manning condensation, parameters $u$ and $f$ should not appear independently but only as the $u f^2$, as will be shown later. This suggests that the problem is 2-parametric, with the second parameter chosen as $r_\mathrm{D} / l_\mathrm{0}$ to quantify how strongly screened Coulomb interactions are.

In this work, we assume that Manning condensation of counterions is neglected, which is justified provided that the distance between the adjacent charges is higher than the solution Bjerrum length
\begin{equation}
    \frac{l_\mathrm{B}}{A} \simeq u f \ll 1
\label{eq:Manning}
\end{equation}
In this respect, both flexible and semiflexible PEs are assumed to have low linear charge density.~\cite{degennes-1976, BJ-review-1996, DR-2005} Theory extension to high charge densities, i.e., to $A \leq l_\mathrm{B}$, appears notoriously difficult because it requires (i) solving the mean-field Poisson-Bolzmann equation in cylindrical geometry to find the fraction of condensed counterions~\cite{deshkovskii-2001, trizac-2016} and (ii) taking into account strong charge correlations between charges on the chain and condensed counterions.~\cite{dobrynin-2006, DR-2006-neck} To the best of our knowledge, both aspects have never been appropriately combined, and the case of non-negligible Manning condensation remains for future work. 

We should, however, note that in the simulations performed in the accompanying paper,~\cite{EPL-part2} counterions and salt are considered implicitly, by using the pairwise interaction potential between the charges given by eq.~\ref{eq:1}. This enables testing and corroborating the scaling laws derived for {\it the theoretical model} outlined in this section, even at high values of the parameters $u$ and $f$, when the Manning condensation would have already taken place for explicit counterions present in the system.

\section{Critical Review of Selected Earlier Theories}
\label{sec:stiffening}

\begin{figure}
\centering
\includegraphics[width=6.5in]{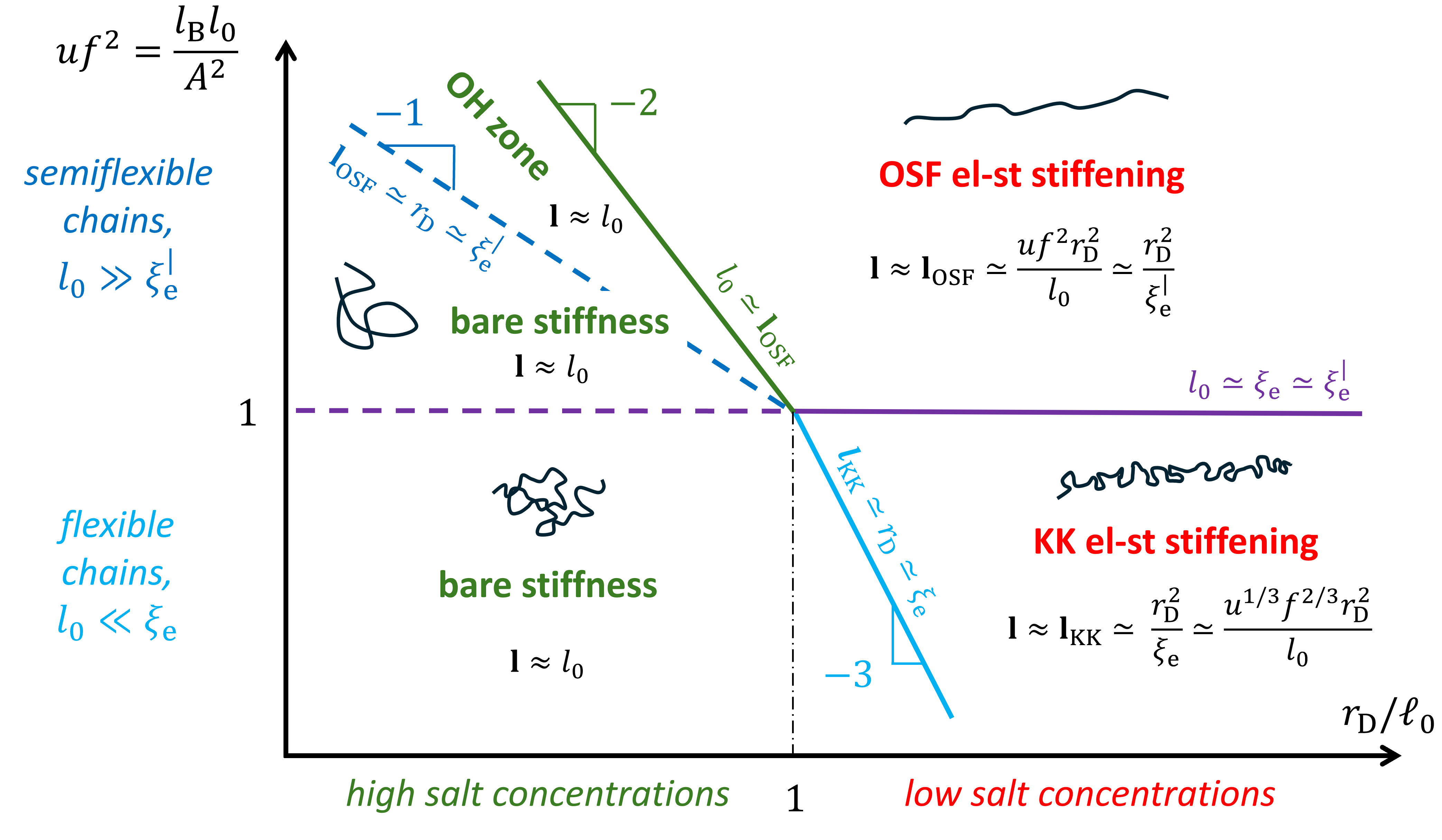}
\caption{Scaling diagram of electrostatic stiffening domains. Boundaries between the domains are demarcated by solid lines, and the flexible/semiflexible crossover is defined by $u f^2 \simeq l_\mathrm{B} l_\mathrm{0} / A^2 \simeq 1$. In bare stiffness domain (green labels), the contribution of electrostatic stiffening is minor, $\textbf{l}_\mathrm{e} \ll \textbf{l} \simeq l_\mathrm{0}$, and negligible at the scaling level of accuracy; the OH zone (green) belongs to the bare stiffness domain. In OSF el-st stiffening domain and KK el-st stiffening domain (red labels), which correspond to semiflexible and flexible PEs, Coulomb interactions dominate chain bare stiffness, $\textbf{l}\simeq \textbf{l}_\mathrm{e} \gg l_\mathrm{0}$.  For clarity, this diagram focuses on differences in {\it local} chain statistics, while the finite-length and excluded-volume effects that determine the global chain statistics and the standard scaling regimes are omitted.}
\label{fig:1}
\end{figure}

In this section, a critical review of earlier theories of electrostatic stiffening is provided in order to construct the diagram of scaling domains shown in Figure~\ref{fig:1}. Here, we adopt the ``domain'' nomenclature to distinguish them from the standard scaling regimes for finite-length chains: as will be shown later, each domain comprises several scaling regimes, i.e., domain $>$ zone $>$ regime. Different domains in the diagram of Figure~\ref{fig:1} are delineated based on the local conformational statistics of the PE, that is, the \textit{local stiffening}. For methodological clarity, the effects of finite chain length $N$ and the effective excluded volume interactions between the stiffened (locally rodlike) quasi-monomers, which control the global (e.g., ideal- versus swollen-coil) conformational statistics and, hence, the $N$-dependence of the chain size, are not considered in this section. They will be superimposed on this diagram later, in Section~\ref{sec:diag}, where we will construct the complete diagram of conformational regimes for finite-length PE chains. 

The diagram of local stiffening in Figure~\ref{fig:1} is constructed in the coordinates of the electrostatic parameter $uf^2 = l_\mathrm{B} l_\mathrm{0} / A^2$ versus the reduced Debye radius $r_\mathrm{D} / l_\mathrm{0}$. Along the $Y$-axis, the diagram is divided by the violet crossover (dashed and then solid line) given by eq.~\ref{cross:stiff} into two parts: the upper and lower parts correspond to semiflexible and flexible chains, respectively. Along the $X$-axis, the high- and low-salt concentration parts are distinguished, with the respective domains labeled in green and red. 

Electrostatic stiffening for semiflexible chains, $u f^2 \gg 1$, has been considered by Odijk~\cite{odijk-1977} and Skolnick and Fixman,~\cite{SF-1977} hence the name ``OSF theory''. The local stiffening occurs at low salt concentrations, in the domain labeled ``OSF el-st stiffening'' in red. The domain of high salt concentrations is labeled in green as the ``bare stiffness'', and the total persistence length is equal to the bare one herein, $\textbf{l} \simeq l_\mathrm{0}$. Within this domain, the separate zone, where stiffening is absent but electrostatic excluded volume interactions remain strong and anisotropic, is distinguished. This zone is called the ``OH zone'' because it was first described by Odijk and Houwaart.~\cite{odijk-1978}

For flexible PEs, $u f^2 \ll 1$, there are also two domains: the high-salt ``bare stiffness'' (green) domain with $\textbf{l}\simeq l_\mathrm{0}$ and the low-salt (red) domain of electrostatic stiffening. Since the stiffening in flexible PEs was first described by Khokhlov and Khachaturian,~\cite{KK-1982} this domain is labeled ``KK el-st stiffening''. In the bare stiffness domain, excluded volume interactions always remain isotropic, and no analog of the OH regime appears for flexible chains. A more detailed analysis of each of the domains is provided below.

\subsection{Odijk-Skolnick-Fixman (OSF) Theory for Semiflexible Polyelectrolytes~\cite{odijk-1977, SF-1977}}

\subsubsection{Bare Stiffness Dominance}

The original OSF calculations were performed within the theoretical model described in Section~\ref{sec:setup} for intrinsically stiff chains.  The resulting electrostatic stiffening, that is, the electrostatic contribution to the total persistence length $\textbf{l}$, is given by~\cite{odijk-1977, SF-1977, GK-book}
\begin{equation}
    \textbf{l}_\mathrm{e} = \textbf{l}_\mathrm{\mathrm{OSF}} \simeq \frac {u f^2 r_\mathrm{D}^2}{l_\mathrm{0}}  
\label{OSF}
\end{equation}
The original OSF calculations apply to the bare stiffness domain because they are perturbative and consider the single-mode bending with respect to the perfectly rodlike conformation. Therefore, the result of eq.~\ref{OSF} remains rigorous at $l_\mathrm{0} \gg \textbf{l}_\mathrm{\mathrm{OSF}}$, i.e., at high salt concentrations only:
\begin{equation}
u f^2  \ll \left( \frac{r_\mathrm{D}} {l_\mathrm{0}} \right)^{-2}
\end{equation}
In Figure~\ref{fig:1}, the respective crossover is shown with the green solid line and captioned as $l_\mathrm{0} \simeq \textbf{l}_\mathrm{\mathrm{OSF}}$. This crossover and the requirement of the high bare stiffness, $u f^2 \gg 1$, which ensures the OSF analysis applicability, delineate the bare stiffness domain for semiflexible PEs in the diagram shown in Figure~\ref{fig:1}. In this domain, PE is locally rodlike and experiences only weak (perturbative) electrostatic stiffening
\begin{equation}
    \textbf{l}= l_\mathrm{0} + \textbf{l}_\mathrm{\mathrm{OSF}} \simeq l_\mathrm{0}
\end{equation}
The last, approximate equality is due to $\textbf{l}_\mathrm{\mathrm{OSF}} \ll l_\mathrm{0}$ in the bare stiffness domain. Note that the OH zone, whose origin is discussed below, also belongs to the bare stiffness domain owing to $\textbf{l}\simeq l_\mathrm{0}$ there.

\subsubsection{Electrostatic Stiffness Dominance}

It was, however, suggested~\cite{odijk-1977, SF-1977} that the OSF result should also hold when it is non-perturbative and $\textbf{l}_\mathrm{\mathrm{OSF}} \gg l_\mathrm{0}$, i.e., at low salt concentrations or equivalently, high Debye radii. In this domain, the total chain stiffness is dominated by the electrostatic contribution
\begin{equation}
    \textbf{l} = l_\mathrm{0} + \textbf{l}_\mathrm{\mathrm{OSF}} \simeq \textbf{l}_\mathrm{\mathrm{OSF}} 
    \simeq \frac {u f^2 r_\mathrm{D}^2}{l_\mathrm{0}} 
\label{OSF-elst-dom}
\end{equation}
The respective domain entitled the ``OSF el-st stiffening'' corresponds to the upper right corner of the diagram shown in Figure~\ref{fig:1}. The violet boundary given by $u f^2 \simeq 1$ demarcates this domain from that of flexible PEs, where chains are no longer locally rodlike, and the OSF analysis is no longer applicable.

It should also be noted that the OSF calculation replaces the pairwise summation over discrete charges with a continuous integration, which is equivalent to assuming that the charge is homogeneously smeared along the chain. This approximation requires the distance between adjacent charges to be much smaller than the Debye radius, $A \ll r_\mathrm{D}$,~\cite{GK-book}  or equivalently $r_\mathrm{D} f / l_\mathrm{0} \gg 1$, and thus defines the range of applicability of the OSF result.

\subsection{Khokhlov-Khachaturian (KK) Theory for Flexible Polyelectrolytes~\cite{KK-1982}}

\subsubsection{Electrostatic Stiffness Dominance}

As discussed in Section~\ref{sec:setup}, PEs are considered (termed) flexible at $u f^2 \ll 1$. At low salt concentrations, PE at intermediate lengths represents the linear array of Gaussian electrostatic blobs,~\cite{degennes-1976, BJ-review-1996, DR-2005, D-2020} resembling stiff PE chains with the effective size of the coarse-grained monomer equal to $\xi_\mathrm{e}$, see Figure~\ref{fig:A-blobs}b. KK suggested that the OSF theory can be extended to flexible PEs by substituting the linear charge density $e f / l_\mathrm{0}$ of the rodlike chain with that of an array of electrostatic blobs, $e f g_\mathrm{e} / \xi_\mathrm{e}$. This is equivalent to the renormalization of the fraction of the charged monomers:
\begin{equation}
    f \quad \to \quad \widetilde{f} \simeq \frac{f g_\mathrm{e} l_\mathrm{0}} {\xi_\mathrm{e}} \simeq u^{-1/3} f^{1/3}
\label{eq:renorm}
\end{equation}
Here, the number $g_\mathrm{e}$ of bare statistical segments in the blob is controlled by local Gaussian statistics of the PE, $g_\mathrm{e} \simeq \left( \xi_\mathrm{e} / l_\mathrm{0} \right)^2$. By combining this $f \to \widetilde{f}$ renormalization with the OSF result given by eq.~\ref{OSF-elst-dom}, KK found that the EPL for flexible chains is given by
\begin{equation}
    \textbf{l} \simeq \textbf{l}_\mathrm{\mathrm{KK}}  \simeq u^{1/3} f^{2/3} \frac{r_\mathrm{D}^2}{l_\mathrm{0}} \simeq \frac{r_\mathrm{D}^2}{\xi_\mathrm{e}}
\label{eq:ell_KK}
\end{equation}
This result has been later rederived using variational calculations~\cite{witten-1995, thirumalai-1999, netz-2004} and supported by simulations,~\cite{everaers-2002, shklovskii-2002} at least in the sense that KK scaling of $\textbf{l}_\mathrm{\mathrm{KK}}  \sim r_\mathrm{D}^2$ works much better in the asymptotic limit of long chains than the linear dependence $\textbf{l}_\mathrm{\mathrm{KK}}  \sim r_\mathrm{D}$ proposed by Barrat and Joanny.~\cite{BJ-1993} The respective domain entitled ``KK el-st stiffening'' corresponds to the bottom right corner of the diagram shown in Figure~\ref{fig:1}. 

Note that the Manning condensation on the array of Gaussian electrostatic blobs~\cite{khokhlov-1980} is negligible when $ u \widetilde{f} \ll 1$, that is, 
\begin{equation}
u f^{1/2} \ll 1
\label{eq:Manning-flex}
\end{equation}
as follows from eq.~\ref{eq:renorm}. This inequality (assumed to be satisfied) also guarantees a consistent definition of the Gaussian electrostatic blob $\xi_\mathrm{e}$ as an entity comprising many charges, $f g_\mathrm{e} \simeq \left( u f^{1/2} \right)^{-2/3} \gg 1$. To summarize, the requirement for the absence of Manning condensation is given by eq.~\ref{eq:Manning} for stiff chains (those with $u f^2 \gg 1$) and eq.~\ref{eq:Manning-flex} for flexible chains (those with $u f^2 \ll 1$). It also deserves mentioning that, for simulations without explicit counterions, Manning condensation is not a concern. It is sufficient to have many monomers (rather than charges) per blob, $u f^{2} \ll 1$ --- a condition that is always satisfied for PEs classified as flexible.

The other limitation stems from the fact that, in the spirit of the OSF calculations, discrete charges are replaced by a continuously smeared charge density. This approximation (summation to integration change) is justified provided the effective distance $ \widetilde{A} = l_\mathrm{0} / \widetilde{f}$ between the charges on the coarse-grained chain of el-st blobs is much lower than the Debye radius, $ \widetilde{A} \simeq \left( A l_\mathrm{0} l_\mathrm{B} \right)^{1/3} \ll r_\mathrm{D}$, or equivalently, $r_\mathrm{D} \widetilde{f} / l_\mathrm{0} \simeq r_\mathrm{D} f^{1/3} / l_\mathrm{0} u^{1/3} \gg 1$.

KK electrostatic stiffening always dominates the bare stiffness of the chain because $\textbf{l}_\mathrm{\mathrm{KK}}  \gg \xi_\mathrm{e} \gg l_\mathrm{0}$. The boundary for the KK domain cannot be found from $\textbf{l}_\mathrm{\mathrm{KK}}  \simeq l_\mathrm{0}$ because $ \widetilde{f}$ is consistently defined only at $\textbf{l}_\mathrm{\mathrm{KK}}  \gg \xi_\mathrm{e}$. Therefore, the applicability of the KK scaling is limited by $\textbf{l}_\mathrm{\mathrm{KK}}  \simeq \xi_\mathrm{e}$ crossover, which, due to eq.~\ref{eq:ell_KK}, corresponds to simultaneous equality between \textit{three} characteristic lengths:
\begin{equation}
    \textbf{l}_\mathrm{\mathrm{KK}}  \simeq r_\mathrm{D} \simeq \xi_\mathrm{e}
\label{triple-cross-KK}
\end{equation}
Physically, when the Debye radius $r_\mathrm{D}$ becomes comparable to or smaller than the el-st blob size $\xi_\mathrm{e}$, the PE chain can no longer be viewed as a rodlike sequence of $\xi_\mathrm{e}$-blobs and begins to bend on this length scale. Crossover given by eq.~\ref{triple-cross-KK} can also be written as 
\begin{equation}
    u f^2 \simeq \left( \frac{r_\mathrm{D}}{l_\mathrm{0}} \right)^{-3} 
\label{triple-cross-KK-2}
\end{equation}
In Figure~\ref{fig:1}, this boundary is shown in solid blue and delineates regions with and without essential electrostatic stiffening for flexible chains.

\subsubsection{Bare Stiffness Dominance}

Beyond this crossover, the Debye radius $r_\mathrm{D}$ is lower than $\xi_\mathrm{e}$, and PE is no longer rodlike at intermediate lengths. Here, Coulomb repulsions become weak and effectively short-range pairwise, so that they can be described using the concept of electrostatic excluded volume,~\cite{KK-1982} as discussed in Section~\ref{sec:EEV}. One can therefore expect the absence of any electrostatic stiffening in this domain:
\begin{equation}
    \textbf{l} \simeq l_\mathrm{0}
\end{equation}

The OSF and KK results are summarized in the diagram of Figure~\ref{fig:1}, which presents scaling domains of different electrostatic stiffening. 
The green solid boundary $l_\mathrm{0} \simeq \textbf{l}_\mathrm{\mathrm{OSF}}$ and the blue solid boundary $\textbf{l}_\mathrm{\mathrm{KK}}  \simeq r_\mathrm{D} \simeq \xi_\mathrm{e}$ completely delineate the regions with/without significant electrostatic stiffening for semiflexible and flexible PEs, respectively. 

However, one may not be completely satisfied with these two boundaries only. Indeed, the KK theory is the direct extension of the OSF theory achieved via $f \to \widetilde{f}$ renormalization. If so, why is the termination of the KK domain given by {\it triple} equality of characteristic lengths, while it is not the case for the OSF domain? If the electrostatic blob is an important length scale for flexible PEs, should there be an {\it analog} of it for semiflexible chains? In the next section, these questions are answered, and one more important crossover, $\textbf{l}_\mathrm{\mathrm{OSF}} \simeq r_\mathrm{D} \simeq \xi_\mathrm{e}^{|}$, shown as the blue dashed line in Figure~\ref{fig:1}, is derived.

\section{Rodlike Electrostatic Blob and Generalized Electrostatic Persistence Length}
\label{sec:elst-blob}

\subsection{Rodlike Electrostatic Blob}

Since the Gaussian ($\Theta$-solvent) el-st blob, $\xi_\mathrm{e}$ given by eq.~\ref{elst-blob}, is an important length scale for the flexible chains, one can expect that the analogous length should also be introduced for semiflexible PEs. The general definition of the el-st blob for chains with the local conformational statistics described by the exponent $\nu$ is equal to
\begin{equation}
    \xi_\mathrm{e}^{(\nu)} \simeq l_\mathrm{0} \left( u f^2 \right)^{ - \nu / (2 - \nu)}
\label{eq:elst-blob}
\end{equation}
This result simply follows from the requirement of the Coulomb energy to be about the thermal energy per blob, $( e f g_\mathrm{e}^{(\nu)} )^2 / \epsilon \xi_\mathrm{e}^{(\nu)} \simeq k_\mathrm{B} T $, and unperturbed chain statistics inside it, $\xi_\mathrm{e}^{(\nu)} \simeq l_\mathrm{0}   \left( g_\mathrm{e}^{(\nu)} \right)^{\nu}$.

\begin{figure}
\centering
\includegraphics[width=3.25in]{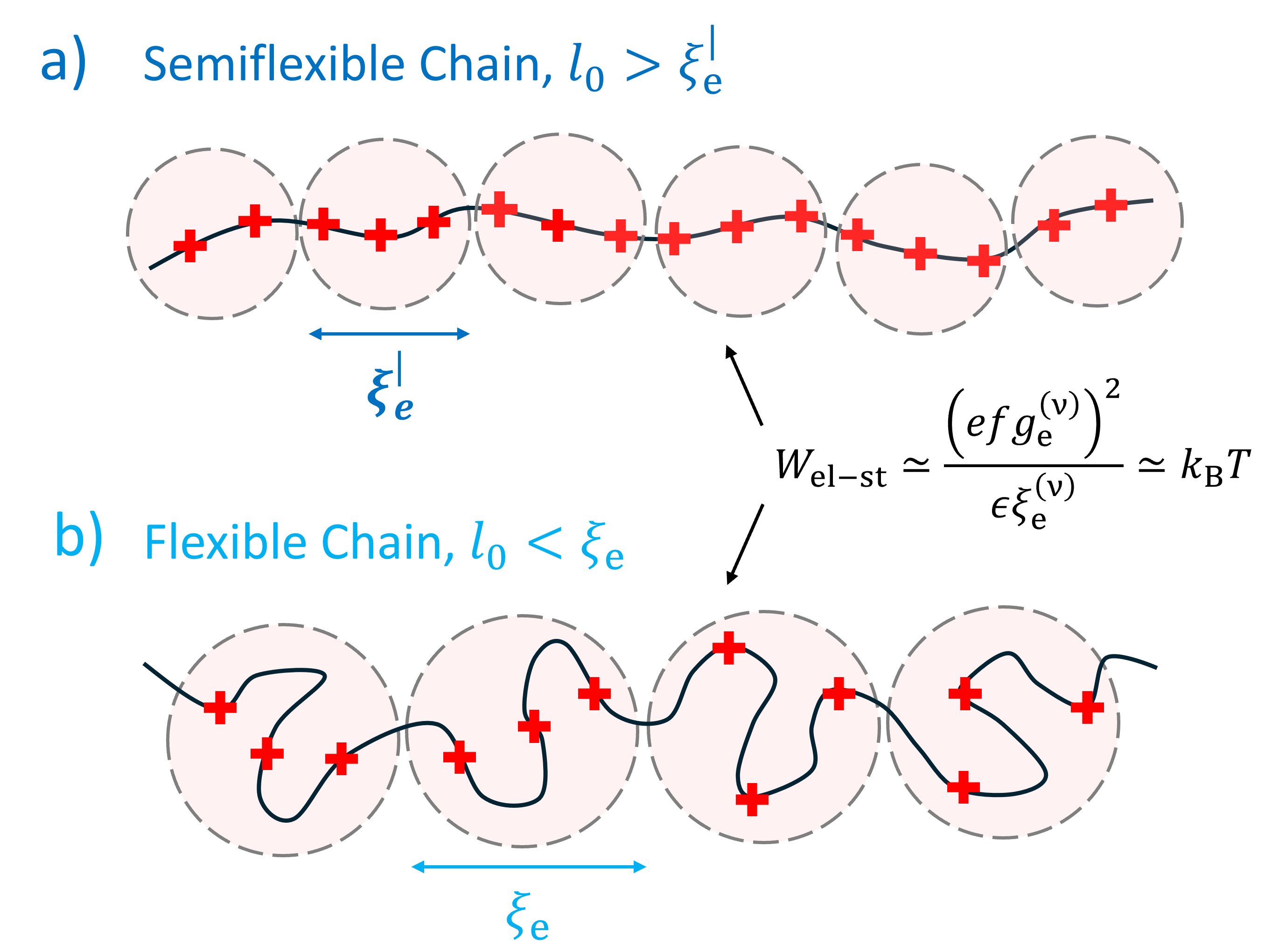}
\caption{Schematic representation of a) semiflexible polyelectrolyte divided into rodlike electrostatic blobs $\xi_\mathrm{e}^{|}$ and b) flexible polyelectrolyte consisting of Gaussian ($\Theta$-solvent) electrostatic blobs $\xi_\mathrm{e}$. Coulomb energy of both blobs is about the thermal energy, $W_\mathrm{el-st} \simeq k_\mathrm{B} T$.}
\label{fig:A-blobs}
\end{figure}

Hence, for rodlike chains with $\nu = 1$, it should be defined as
\begin{equation}
    \xi_\mathrm{e}^{|} \simeq l_\mathrm{0} \left( u f^2 \right)^{-1} = \frac{A^2}{l_\mathrm{B}}
\label{eq:xi-rod-def}
\end{equation}
The physical meaning of this result is straightforward. The free energy of Coulomb repulsions of the PE charges within the el-st blob is on the order of the thermal energy, $k_\mathrm{B} T$. The difference compared to the well-known cases of $\Theta$ and good solvent, $\nu = 1/2$ and $\nu \approx 3/5$, is that the chain does not change its local statistics to rodlike at $r > \xi_\mathrm{e}^{|}$ because, for stiff chains, it is already rodlike, even inside the el-st blob, at $r < \xi_\mathrm{e}^{|}$. This is shown in Figure~\ref{fig:A-blobs}a.

The rodlike el-st blob is also associated with the Coulomb cost of local bending fluctuations. One may argue that having a change in the chain orientation by $ \delta \theta \simeq 1$ inside it, i.e., $\mathbf{u}(s) \cdot \mathbf{u} (s+\xi_\mathrm{e}^{|}) \simeq 1/2$, results in the extra Coulomb energy on the order of the thermal energy $k_\mathrm{B} T$. (Recall that $\mathbf{u}(s) \cdot \mathbf{u} (s+\xi_\mathrm{e}^{|}) = 1$ and $ \delta \theta = 0$ for the perfectly straight rod, where $\mathbf{u}(s)$ denotes the unit vector along the chain direction.~\cite{GK-book}) 

For semiflexible chains with $u f^2 \gg 1$, the rodlike electrostatic blob is smaller than the Kuhn segment, $\xi_\mathrm{e}^{|} \ll l_\mathrm{0}$. However, this blob corresponds to a high number of chemical monomers and carries multiple charges 
\begin{equation}
\frac {q_\mathrm{e}^{|}} {e} \simeq \frac{ f \xi_\mathrm{e} } { l_\mathrm{0} } \simeq \left( uf \right)^{-1} \gg 1, 
\end{equation}
with the last inequality being satisfied due to the absence of Manning condensation, cf. eq.~\ref{eq:Manning}. At that, $g_\mathrm{e}^{|} \simeq \xi_\mathrm{e}^{|} / l_\mathrm{0} \simeq (uf^2)^{-1} \ll 1$ should be considered the {\it fraction} of the Kuhn segment within one rodlike el-st blob. 

The $\xi_\mathrm{e}^{|}$ definition provided by eq.~\ref{eq:xi-rod-def} immediately enables writing down the OSF result as
\begin{equation}
    \textbf{l}_\mathrm{\mathrm{OSF}} \simeq \frac{r_\mathrm{D}^2} {\xi_\mathrm{e}^{|}}, 
\end{equation}
which makes it completely analogous to the KK results for flexible chains, $\textbf{l}_\mathrm{\mathrm{KK}}  \simeq r_\mathrm{D}^2 / \xi_\mathrm{e}$, cf. eq.~\ref{eq:ell_KK}.

In a similar vein, it was earlier suggested~\cite{netz-2001, dobrynin-2009, rubinstein-2017} that the charged semiflexible PE can be considered a neutral wormlike chain under stretching; variational calculations by Dobrynin et al. yielded the respective stretching force.~\cite{dobrynin-2009} After $\xi_\mathrm{e}^{|}$ is introduced, this force in the salt-free case can be written as the inverse tension blob size, with the latter equal to the rodlike el-st blob size:
\begin{equation}
\frac { \text{force} } {k_\mathrm{B} T} \simeq \frac {u f^2} {l_\mathrm{0}} 
\simeq \frac{1} {\xi_\mathrm{e}^{|}}  
\label{eq:force}
\end{equation}
This result represents a direct extension of the argument from flexible to semiflexible PEs: two electrostatic blobs repel each other with the $k_\mathrm{B} T$ energy, generating $k_\mathrm{B} T / \xi_\mathrm{e}^{(\nu)}$ force. In flexible PEs, this Coulombic stretching changes the global $\nu$ value to 1, whereas for semiflexible PEs it does not because the local chain statistics are already rodlike, $\nu = 1$. 

Nevertheless, $\xi_\mathrm{e}^{|}$ is important for identifying scaling crossovers relevant to the EPL problem. The crossover between flexible and semiflexible PEs, given by $u f^2 \simeq 1$ and shown with violet line in Figure~\ref{fig:1}, can also be expressed as the equality between the Gaussian and rodlike el-st blob sizes; moreover, at this crossover, they both are also equal to the bare persistence length:
\begin{equation}
    \xi_\mathrm{e} \simeq \xi_\mathrm{e}^{|} \simeq l_\mathrm{0} 
\label{eq:stiff/flex} 
\end{equation}
This reflects the change in the intra-blob conformations from locally ideal-coil with $\nu = 1/2$ to rodlike with $\nu = 1$, as illustrated in Figure~\ref{fig:A-blobs}.

Let us continue drawing parallels and transferring results from flexible to semiflexible PEs, i.e., from the KK back to the OSF regime. In the former case, the triple crossover $\textbf{l}_\mathrm{\mathrm{KK}}  \simeq r_\mathrm{D} \simeq \xi_\mathrm{e}$ given by eq.~\ref{triple-cross-KK} defines the point where Coulomb interactions cease to stiffen the chain, see blue solid line in Figure~\ref{fig:1}. For semiflexible chains, formally similar boundary reads
\begin{equation}
\textbf{l}_\mathrm{\mathrm{OSF}} \simeq r_\mathrm{D} \simeq \xi_\mathrm{e}^{|}   
\end{equation}
or equivalently
\begin{equation}
    u f^2 \simeq \left( \frac{l_\mathrm{0}} {r_\mathrm{D}} \right)^{-1}
\label{eq:cross-OSF-blobs}
\end{equation}
and is shown with the blue dashed line in Figure~\ref{fig:1}. Beyond this crossover, Coulomb interactions within the rodlike el-st blob become substantially screened. Thus, the addition of salt can be considered as decreasing the force that effectively stretches the chain, which for $r_\mathrm{D} < \xi_\mathrm{e}^{|}$ no longer obeys eq.~\ref{eq:force} because of the salt screening of inter-blob Coulomb repulsions. In contrast to the case of flexible PEs, this does not change the scaling of the total persistence length, $\textbf{l}\simeq l_\mathrm{0} \gg \textbf{l}_\mathrm{\mathrm{OSF}}$, i.e., the bare stiffness dominance persists on both sides of this crossover owing to the high $l_\mathrm{0}$ values for semiflexible chains. However, this crossover corresponds to a change in the nature (strength and effective anisotropy) of Coulomb interactions and, hence, in the chain structure and the power law for its size, as demonstrated below. For this reason, this crossover delineating the OH zone within the bare stiffness domain is included in Figure~\ref{fig:1}.

\subsection{Generalized Electrostatic Persistence Length}

The el-st blob can be considered a monomer of the coarse-grained PE chain, which interacts with neighboring monomers with the Coulomb energy of order $k_\mathrm{B} T$. The bending stiffness of the PE for an arbitrary $\nu$ depends only on the Debye radius $r_\mathrm{D}$ and the generalized electrostatic blob size $\xi_\mathrm{e}^{(\nu)}$:
\begin{equation}
    \textbf{l}_\mathrm{e}^{(\nu)} \simeq \frac{r_\mathrm{D}^2} {\xi_\mathrm{e}^{(\nu)}}
\label{EPL-gen}
\end{equation}
This result can be proven by (i) coarse-graining the chain to el-st blob size monomers and (ii) in the spirit of KK renormalization given by eq.~\ref{eq:renorm}, defining the effective linear charge density as $\widetilde{f} \simeq e f g_\mathrm{e}^{(\nu)} / \xi_\mathrm{e}^{(\nu)}$. By substituting $f \simeq \widetilde{f}$ into the OSF result, eq.~\ref{OSF}, and using the blob definition clause, $l_\mathrm{0} u f^2 (g_\mathrm{e}^{(\nu)})^2 / \xi_\mathrm{e}^{(\nu)} \simeq 1 $, one arrives at eq.~\ref{EPL-gen} defining the general EPL of the chain with the arbitrary local exponent $\nu$. For $\nu = 1/2$ and $\nu = 1$ this result reduces to $\textbf{l}_\mathrm{e}^{(\nu = 1/2)} \simeq \textbf{l}_\mathrm{\mathrm{KK}} $ and $\textbf{l}_\mathrm{e}^{(\nu = 1)} \simeq \textbf{l}_\mathrm{\mathrm{OSF}}$, respectively.~\cite{footnote-1} 

Eqs.~\ref{eq:elst-blob} and~\ref{EPL-gen} illustrate why the parameters $u$ and $f$ in the EPL problem appear not independently but through the product $u f^2$ , provided that $r_\mathrm{D} / l_\mathrm{0}$ is chosen as the other independent parameter. The physical interpretation is that, effectively, the charge can be considered homogeneously smeared along the chain of blobs. Therefore, the simultaneous increase in the number of charges (controlled by $f$) and the decrease in the strength of their interactions (controlled by $u$) does not affect the system behavior as long as $uf^2$ remains unchanged. 

\section{Electrostatic Excluded Volume for Non-Adjacent \\ Monomers}
\label{sec:EEV}

From this section onward, we consider nonlocal excluded-volume interactions between non-adjacent monomers. To theoretically describe chain conformational statistics at high salt concentrations (green domain in Figure~\ref{fig:1}), it is useful to apply the concept of electrostatic excluded volume (eev). The concept suggests that, at high salt concentrations, Coulomb interactions between monomers can be treated as short-range pairwise, described by the respective second virial coefficient $B_\mathrm{eev}$. In the present section, careful consideration of this concept is provided to better reveal the range of its applicability and the (often implicit) assumptions invoked. 

\subsection{Naive Approach}

Consider two charges interacting via the pairwise screened Coulomb potential $u_\mathrm{12} (r)$, eq.~\ref{eq:1}. The second virial coefficient of their interaction equals
\begin{equation}
    B_\mathrm{cc} = \frac{1}{2} \int \left[ 1 - \exp \left( - \frac {u_\mathrm{12} (r)} {k_\mathrm{B} T}   \right)  \right] d^3 r
\label{eq:virial}
\end{equation}
To estimate this integral, one can notice that repulsions are strong at short distances and weak at large distances. The square braces in eq.~\ref{eq:virial}, which are equal to the Mayer $f$-function taken with the negative sign, are approximately equal to
\begin{equation}
     1 - \exp \left( - \frac{l_\mathrm{B}}{r} e^{-r/r_\mathrm{D}}  \right) \approx
     \begin{cases}
     1,         & r \leq l_\mathrm{B} \\
     l_\mathrm{B} / r,   & l_\mathrm{B} \leq r \leq r_\mathrm{D} \\
     l_\mathrm{B} e^{-r/r_\mathrm{D}} / r,  & r_\mathrm{D} \leq r 
     \end{cases}
\end{equation}
Here we assumed that $l_\mathrm{B} \ll r_\mathrm{D}$ because this is the necessary condition for the applicability of the linearized Debye-H\"{u}ckel (that is, the RPA) theory of screening,~\cite{LL-book} which provides the functional form for $u_\mathrm{12} (r)$ given by eq.~\ref{eq:1}. The exponent in the Mayer function was linearized when appropriate, at $r \geq l_\mathrm{B} $. By dividing the integral~\ref{eq:virial} into the three respective parts and dropping all the prefactors, one can see that the leading contribution comes from $r \simeq r_\mathrm{D}$ distances between the charges, where their interactions are not yet exponentially weak: 
\begin{equation}
    B_\mathrm{cc} \simeq \int_\mathrm{0}^{l_\mathrm{B}} r^2 dr +
    \int_\mathrm{l_\mathrm{B}}^{r_\mathrm{D}} \frac{l_\mathrm{B}}{r} r^2 dr +
    \int_\mathrm{r_\mathrm{D}}^{\infty} \frac{l_\mathrm{B}}{r} e^{-r/r_\mathrm{D}} r^2 dr  
    \simeq  l_\mathrm{B}^3 + l_\mathrm{B} r_\mathrm{D}^2 + l_\mathrm{B} r_\mathrm{D}^2 \simeq l_\mathrm{B} r_\mathrm{D}^2 
\label{eq:B-integral}
\end{equation}
Again, here we made use of $l_\mathrm{B} \ll r_\mathrm{D}$. This result represents the second virial coefficient of interaction between two charges, hence the cc subscript. If rescaled to the effective interactions between all the monomers, the concentration of which is $f^{-1}$ times higher, it is equal to
\begin{equation}
    B_\mathrm{eev} \simeq f^2 B_\mathrm{cc} \simeq l_\mathrm{B} f^2 r_\mathrm{D}^2 \simeq l_\mathrm{0} u f^2 r_\mathrm{D}^2
\label{eq:EEV-deriv}
\end{equation}
Another way to derive this well-known result is to consider PE gel in the presence of salt and take into account the Donnan equilibrium between its interior and the outer solution.~\cite{DR-2005}

It should be noted that the final results given by eqs.~\ref{eq:B-integral}–\ref{eq:EEV-deriv} hold even when $l_\mathrm{B} > r_\mathrm{D}$, provided that charges continue interacting via a screened Coulomb potential. Indeed, in this case, $u_\mathrm{12} (r) \simeq l_\mathrm{B} / r \gg 1$ for $r < r_\mathrm{D}$ and $u_\mathrm{12} (r) \simeq l_\mathrm{B} e^{-r/r_\mathrm{D}} / r \ll 1$ for $r > r_\mathrm{D}$, leading to
\begin{equation}
    B_\mathrm{cc} \simeq \int_\mathrm{0}^{r_\mathrm{D}} r^2 dr + \int_\mathrm{r_\mathrm{D}}^{\infty} 
    \frac{l_\mathrm{B}} {r} e^{-r/r_\mathrm{D}}  r^2 dr \simeq r_\mathrm{D}^3 + l_\mathrm{B} r_\mathrm{D}^2 \simeq l_\mathrm{B} r_\mathrm{D}^2  
\end{equation}
This situation, albeit somewhat physically inconsistent within the DH/RPA screening framework, may help test scaling laws in simulations \textit{without explicit counterions} (e.g., using a screened Coulomb potential for pairwise interactions) in regimes where the electrostatic excluded volume concept is relevant.

The problem with the derivation outlined is that it completely neglects the {\it charge connectivity}, which limits the range of its applicability to weak Coulomb interactions only. Eq.~\ref{eq:EEV-deriv} is valid at $\xi_\mathrm{e} < r_\mathrm{D}$ for flexible and at $\xi_\mathrm{e}^{|} < r_\mathrm{D}$ for semiflexible PEs, that is, only in the part of the bare stiffness domain. To extend the naive approach, an alternative method is provided below that can also be used to describe the OH zone.

\subsection{Debye Blobs Approach}

\begin{figure}
\centering
\includegraphics[width=6.5in]{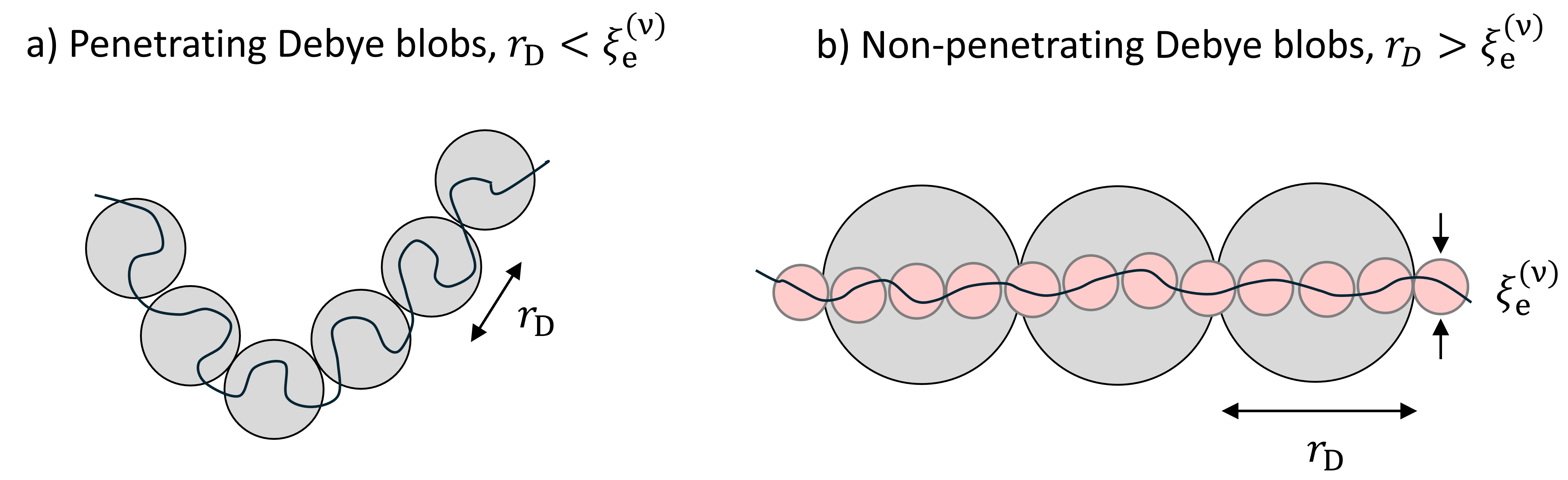}
\caption{Schematic representation of polyelectrolyte chain fragments in the cases of (a) penetrating and (b) non-penetrating Debye blobs, which correspond to weak, $r_\mathrm{D} < \xi_\mathrm{e}^{(\nu)}$, and strong, $r_\mathrm{D} > \xi_\mathrm{e}^{(\nu)}$, electrostatic interactions between them. In the former case, the electrostatic blob (Gaussian or rodlike, depending on chain flexibility) is no longer the relevant length scale for defining local conformational statistics. Debye blobs are shown in gray, and electrostatic blobs in pink. }
\label{fig:B-Debye}
\end{figure}

The Debye blob is simply a fragment of the chain of size about $r_\mathrm{D}$, as shown in Figure~\ref{fig:B-Debye}. Charged monomers residing within the same Debye blob interact with other proximate charges collectively, that is, altogether via an unscreened potential. The key idea is to reduce the interactions between all charges in the PE to interactions between Debye blobs.

Consider PE chains with local $\nu$-statistics that approach each other to within a distance $r_\mathrm{D}$, for example, in a perpendicular orientation. It is the fragments of these chains of size $r_\mathrm{D}$ --- the Debye blobs --- that repel each other via unscreened Coulomb interactions, while repulsions between more distant neighbors are exponentially weak and negligible. The pairwise interaction potential $u_\mathrm{12}^{D} (r)$ between two Debye blobs, each with charge $q_\mathrm{D}$, can be approximated as follows:
\begin{equation}
    u_\mathrm{12}^{D} (r) \simeq 
    \begin{cases}
     q_\mathrm{D}^2 / \epsilon r_\mathrm{D} \equiv W_\mathrm{D},   &   r \leq r_\mathrm{D}  \\
    q_\mathrm{D}^2 e^{-r/r_\mathrm{D}} / \epsilon r \approx 0,    &   r > r_\mathrm{D}
    \end{cases}
\label{eq:pairwise-u-DD}
\end{equation}
At short distances, the strength of repulsion between close/overlapping Debye blobs is independent of the distance between their centers, so $r$ in the denominator is replaced by $r_\mathrm{D}$. At large distances, interactions are exponentially weak and can be assumed to be zero, and dropping this term would not affect the resulting scaling of the second virial coefficient. Similarly, the third term could have been dropped in eq.~\ref{eq:B-integral}. Physically, this suggests treating electrostatic interactions between the Debye blobs as effectively short-range, with a range of order $r_\mathrm{D}$.

The Debye blob charge is defined by the local conformational statistics, $q_\mathrm{D} \simeq e f g_\mathrm{D} \simeq e f \left( r_\mathrm{D} / l_\mathrm{0} \right)^{1 / \nu}$, with $\nu = 1$ for semiflexible (locally rodlike) chains and $\nu = 1/2$ for flexible chains. The resulting strength of inter-blob repulsions upon their pairwise contact is 
\begin{equation}
    \frac {W_\mathrm{D}} {k_\mathrm{B} T} \simeq \frac{q_\mathrm{D}^2} {\epsilon k_\mathrm{B} T r_\mathrm{D}} 
    \simeq u f^2 \left( \frac{r_\mathrm{D}} {l_\mathrm{0}} \right)^{ (2 - \nu) / \nu }
\end{equation}
and the second virial coefficient of interactions between the Debye blobs reduces to
\begin{equation}
    B_\mathrm{DD} = \frac{1}{2} \int \left[ 1 - \exp \left( - \frac {u_\mathrm{12}^\mathrm{D} (r)} {k_\mathrm{B} T}   \right)  \right] d^3 r
    \simeq r_\mathrm{D}^3 \left[ 1 - \exp \left( - \frac{W_\mathrm{D}}{k_\mathrm{B} T}  \right) \right]  
\label{eq:virial-DD}
\end{equation}
Note that here and in eq.~\ref{eq:pairwise-u-DD}, the logarithmic prefactors have been neglected; their calculation can be found in the Appendix.

\subsubsection{Weak Coulomb Interactions: Penetrating Debye Blobs}

We first consider weak Coulomb repulsions between the Debye blobs, $W_\mathrm{D} \ll k_\mathrm{B} T$, presented in Figure~\ref{fig:B-Debye}a. In this case, linearizing the exponent in eq.~\ref{eq:virial-DD} leads to $B_\mathrm{DD} \simeq r_\mathrm{D}^3 W_\mathrm{D} / k_\mathrm{B} T$ for the blob-blob interactions. The resulting effective second virial coefficient for the monomer-monomer interactions is the ratio between $B_\mathrm{DD}$ and the squared number of monomers within the Debye blob $g_\mathrm{D}^2$:
\begin{equation}
    B_\mathrm{eev} \simeq \frac{B_\mathrm{DD}} {g_\mathrm{D}^2} \simeq l_\mathrm{0} u f^2 r_\mathrm{D}^2 
\end{equation}
This result agrees with eq.~\ref{eq:EEV-deriv} obtained within the naive approach.

The requirement of $W_\mathrm{D} \ll k_\mathrm{B} T$ is fulfilled at sufficiently low Debye radii not exceeding the size of the general electrostatic blob $\xi_\mathrm{e}^{(\nu)}$ defined for arbitrary $\nu$:
\begin{equation}
    r_\mathrm{D} < \xi_\mathrm{e}^{(\nu)} \simeq l_\mathrm{0} \left( u f^2 \right)^{ - \nu / (2 - \nu)}
    \simeq
    \begin{cases}
        l_\mathrm{0} \left( u f^2 \right)^{-1/3} \simeq \xi_\mathrm{e}, 
        & \text{for} \quad \nu = 1/2 \\
        l_\mathrm{0} \left( u f^2 \right)^{-1} \simeq \xi_\mathrm{e}^{|}, \
        & \text{for} \quad \nu = 1 
    \end{cases}
\label{eq:lin-EEV-criterion}
\end{equation}
This further clarifies why the crossovers $r_\mathrm{D} \simeq \xi_\mathrm{e}$ and $r_\mathrm{D} \simeq \xi_\mathrm{e}^{|}$ given by eqs.~\ref{triple-cross-KK-2} and \ref{eq:cross-OSF-blobs}, which are shown as blue solid and blue dashed lines in Figure~\ref{fig:1}, are similar. For $r_\mathrm{D} \leq \xi_\mathrm{e}^{(\nu)}$, Coulomb interactions for both flexible and semiflexible chains can be considered within the naive electrostatic excluded volume framework, with the same scaling laws (swollen-coil regime SC) in both cases, as elaborated in Section~\ref{sec:diag}. 

The advantage of the Debye blob approach is the ability to control the validity of the derivations. It shows that the standard electrostatic excluded volume method with $B_\mathrm{eev} \simeq l_\mathrm{0} u f^2 r_\mathrm{D}$ is only valid when Coulomb interactions are {\it weak and linearizable}, that is, when the Debye radius is smaller than the generalized el-st blob, eq.~\ref{eq:lin-EEV-criterion}.

\subsubsection{Strong Coulomb Interactions: Non-penetrating Debye Blobs}

When electrostatic repulsions within the Debye blob are strong, $W_\mathrm{DD} \gg k_\mathrm{B} T$, the exponent in eq.~\ref{eq:virial-DD} is much lower than unity, and the second virial coefficient for the Debye blobs (up to logarithmic corrections derived in the Appendix) is of order $B_\mathrm{DD} \simeq r_\mathrm{D}^3$. This result suggests that Debye blobs do not penetrate each other, see Figure~\ref{fig:B-Debye}b. If recalculated on a per-monomer basis, the electrostatic second virial coefficient is given by
\begin{equation}
    \widetilde{B}_\mathrm{eev}^\mathrm{S} \simeq \frac{r_\mathrm{D}^3} {g_\mathrm{D}^2} \simeq l_\mathrm{0}^{3} \left( \frac{r_\mathrm{D}} {l_\mathrm{0}} \right)^{(3 \nu - 2) / \nu}
\label{eq:B_eev_S}
\end{equation}
This result does not coincide with the $B_\mathrm{eev}$ derived earlier for weak Coulomb interactions; it is therefore denoted $\widetilde{B}_\mathrm{eev}^\mathrm{S}$ to emphasize its validity only for strong repulsions between blobs. Also, $W_\mathrm{DD} \gg k_\mathrm{B} T$ implies that $r_\mathrm{D} > \xi_\mathrm{e}^{(\nu)}$, and each Debye blob comprises several el-st blobs, Gaussian and rodlike for flexible and semiflexible PEs, respectively. El-st blobs repel each other via unscreened Coulomb forces, so for any $\nu < 1$ this would compromise the clause of unperturbed chain statistics within the Debye blob, $g_\mathrm{D} \simeq (r_\mathrm{D} / l_\mathrm{0})^{1/\nu}$. However, for $\nu = 1$ this result remains valid and, as we demonstrate below, can be used to describe chain conformations in the OH regime by using the respective electrostatic excluded volume (second virial coefficient): 
\begin{equation}
    B_\mathrm{OH} \simeq \left. \widetilde{B}_\mathrm{eev}^\mathrm{S} \right|_\mathrm{\nu = 1} \simeq l_\mathrm{0}^2 r_\mathrm{D} 
\label{eq:B_OH}
\end{equation}

The physical meaning of this result is as follows. In the OH regime, in the absence of electrostatic stiffening, the Kuhn segment size is equal to $l_\mathrm{0}$. Owing to Coulomb repulsions, segments do not penetrate each other (because each comprises several Debye blobs, and provided that $A < r_\mathrm{D}$), and their effective thickness is about the Debye radius $r_\mathrm{D}$. Since in the OH regime $ r_\mathrm{D} \ll \textbf{l}_\mathrm{\mathrm{OSF}} \ll l_\mathrm{0}$, each Kuhn segment is effectively a long cylinder of $l_\mathrm{0}$ length and $r_\mathrm{D}$ diameter, which leads to the second virial coefficient of interaction of such cylinders~\cite{GK-book} equal to $B \simeq l_\mathrm{0}^2 r_\mathrm{D}$. 

It should be noted that the result of eq.~\ref{eq:B_OH} cannot be directly applied to the OSF el-st stiffening domain because $B_\mathrm{OH}$ is calculated for the monomer of length $l_\mathrm{0}$, while the OSF Kuhn segment far exceeds this value, $\textbf{l}_\mathrm{\mathrm{OSF}} \gg l_\mathrm{0}$. The problem can be addressed by using the quasi-monomer renormalization with the choice of the monomer size $l_\mathrm{0}$, the length of the string connecting two monomers equal to
\begin{equation}
a \simeq (l_\mathrm{0} \textbf{l}_\mathrm{\mathrm{OSF}})^{1/2}    
\label{eq:a-QM2}
\end{equation}
and the corresponding $B_\mathrm{OH} \simeq l_\mathrm{0}^2 r_\mathrm{D}$ for pairwise interactions of these new monomers. (This corresponds to the chain division of type 1 in the notation of $\S$13 of ref.~\citenum{GK-book}). However, it is easier to choose the quasi-monomers to be the stiffened OSF segments $\textbf{l}_\mathrm{\mathrm{OSF}}$ and calculate the second virial coefficient for them (division of type 2) --- the approach used in the remainder of the paper.

\section{Diagram of Conformational Regimes}
\label{sec:diag}

\begin{figure}
\centering
\includegraphics[width=6.5in]{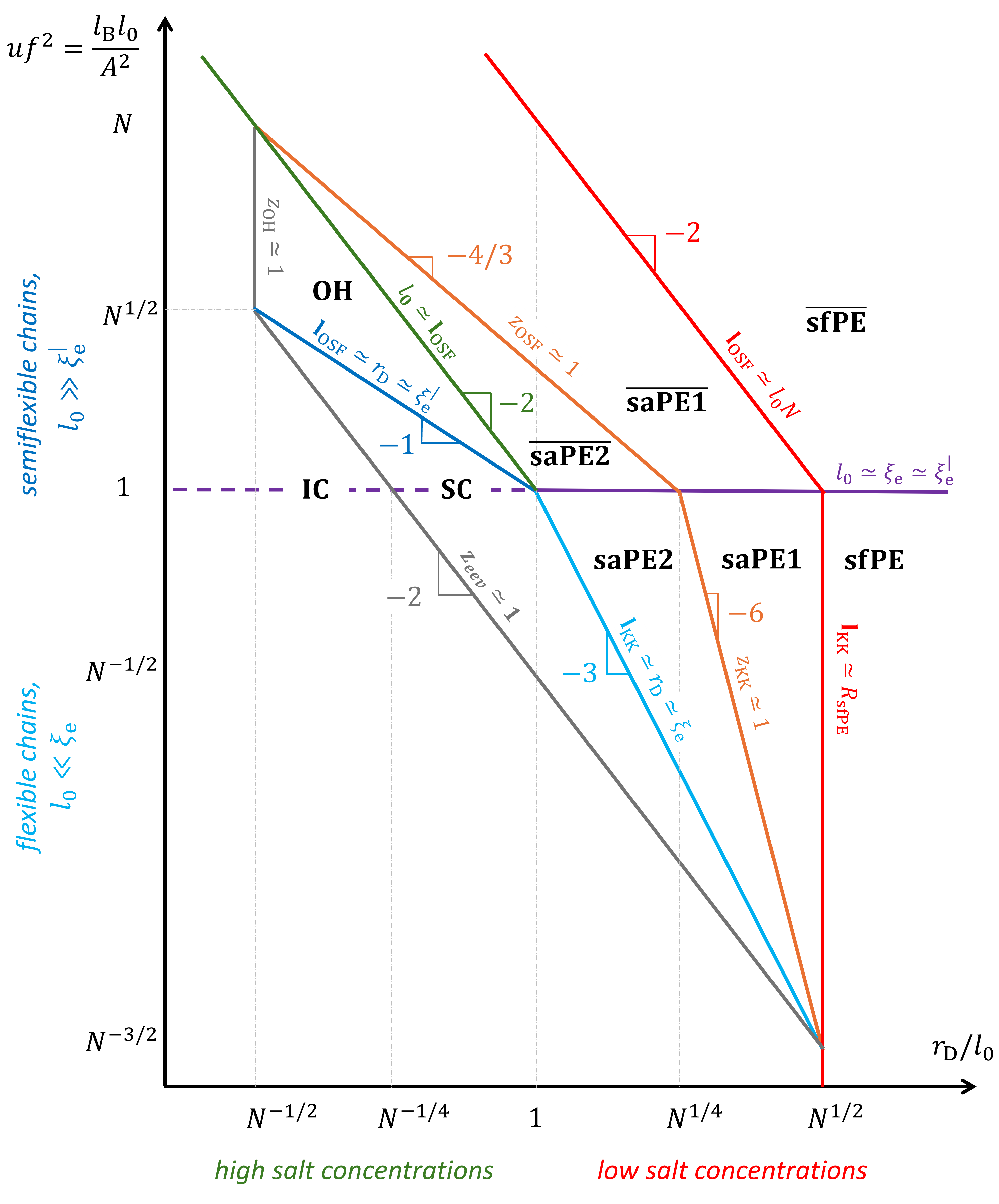}
\caption{Scaling diagram of conformational regimes for single-chain finite-length polyelectrolytes in the salt-added solution with the Debye screening radius $r_\mathrm{D}$. Chains consist of $N$ bare Kuhn segments, each of $l_\mathrm{0}$ length. The respective power laws for the polyelectrolyte dimensions are summarized in Table~\ref{table:1}. }
\label{fig:2}
\end{figure}

We are now in a position to provide a scaling picture of the PE chain and the respective scaling laws for the chain size in each of the regimes. The diagram of the chain stiffening constructed in Figure~\ref{fig:1} demonstrates that the domain of flexible chains, $u f^2 \ll 1$, contains a lower number of crossovers. For this reason, we start our analysis with it. In the next subsection, the results are generalized to semiflexible PEs with $u f^2 \gg 1$. The resulting scaling diagram of the PE conformational regimes is shown in Figure~\ref{fig:2}.

\begin{figure}
\centering
\includegraphics[width=6.5in]{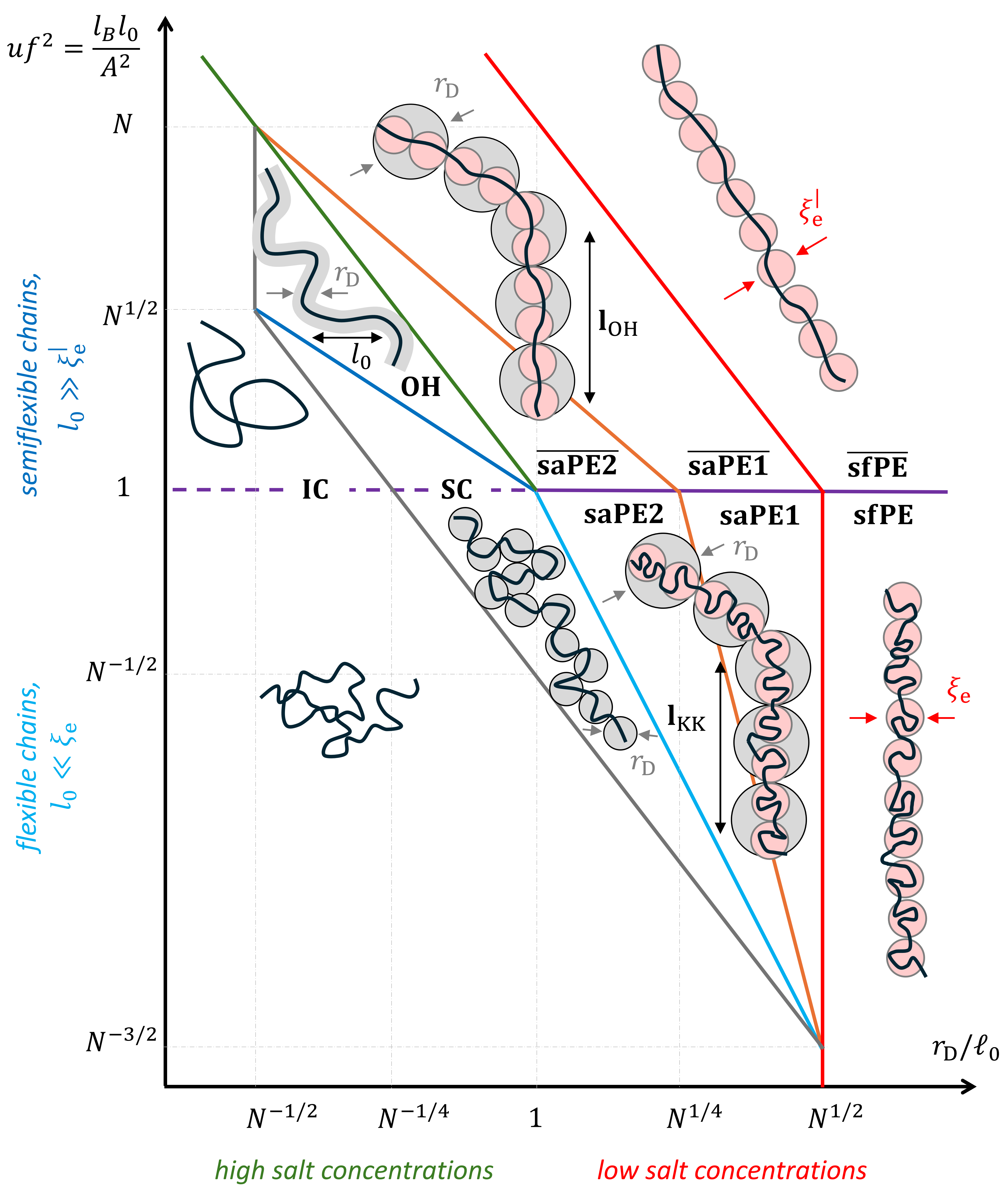}
\caption{The same scaling diagram of conformational regimes of single-chain polyelectrolyte in the salt-added solution as in Figure~\ref{fig:2} with the schematic illustration of the chain conformations. The electrostatic blobs $\xi_\mathrm{e}$ and $\xi_\mathrm{e}^{|}$ are shown as pink circles, and the Debye blobs are shown as gray circles.}
\label{fig:2pic}
\end{figure}

\subsection{Flexible Polyelectrolytes~\cite{KK-1982}, $u f^2 \ll 1$}

We follow the KK work~\cite{KK-1982} and re-derive the scaling regimes first reported therein. In the regime notation, we follow our earlier work, ref.~\citenum{RGJ-2024}, and denote sfPE and saPE the regimes of essentially salt-free and salt-added PEs, respectively.

\subsubsection{Salt-Free PE with Rodlike Global Statistics (Regime sfPE)}

The scaling picture of the salt-free flexible PE is due to de Gennes et al.~\cite{degennes-1976}, who suggested considering the chain as a linear array (stretch) of the electrostatic blobs of the length
\begin{equation}
    R_\mathrm{sfPE} \simeq L_\mathrm{e} \simeq \xi_\mathrm{e} \frac{N} {g_\mathrm{e}} 
    \simeq \frac{l_\mathrm{0}^2 } {\xi_\mathrm{e}} N
    \simeq l_\mathrm{0} u^{1/3} f^{2/3} N
\label{eq:sfPE}
\end{equation}
Note that here and below, all logarithmic corrections to the chain size have been neglected;~\cite{degennes-1976, DR-2005} their values can be found in the Appendix.

\subsubsection{Salt-Added PE with Ideal-Coil Global Statistics
and Local Electrostatic Stiffening (Regime saPE1)}

The EPL for flexible PEs is given by the KK result, eq.~\ref{eq:ell_KK}.
Salt screening begins to manifest itself when the contour length of the chain of electrostatic blobs --- equal to the chain end-to-end distance in the sfPE regime, $L_\mathrm{e} \simeq R_\mathrm{sfPE}$ --- exceeds the EPL. The crossover takes place at $R_\mathrm{sfPE} \simeq \textbf{l}_\mathrm{\mathrm{KK}} $, which can be also expressed as
\begin{equation}
    \left( \frac{r_\mathrm{D}} {l_\mathrm{0}} \right)_\mathrm{sfPE/saPE1} \simeq N^{1/2}
\label{eq:cross-sfPE/saPE1}
\end{equation}
It is shown with the red vertical solid line in the bottom right part of Figure~\ref{fig:2}.

\begin{figure}
\centering
\includegraphics[width=3.25in]{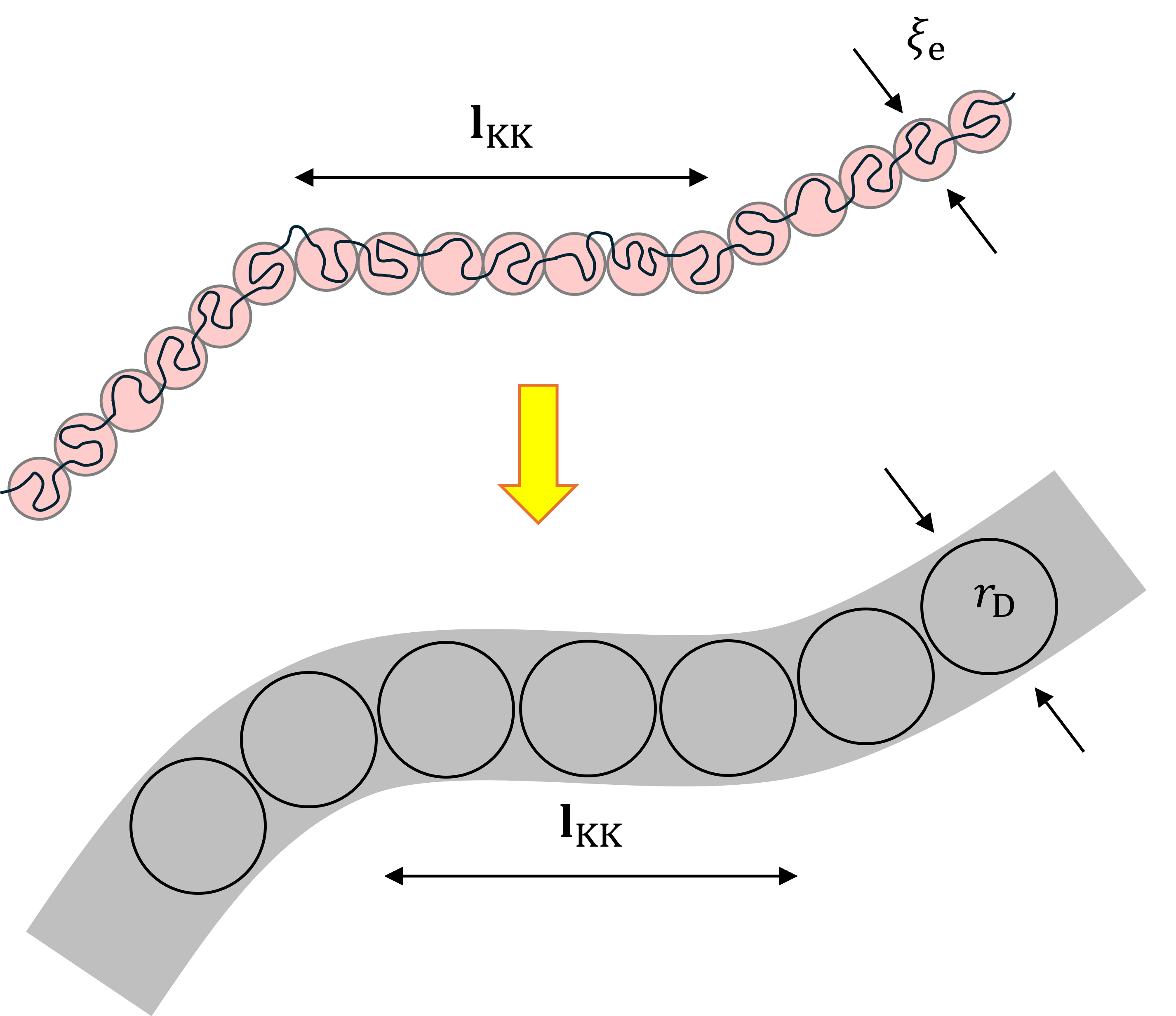}
\caption{Equivalent representation of polyelectrolyte chain in saPE1 and saPE2 regimes as the chain of quasi-monomers with the length $\textbf{l}_\mathrm{\mathrm{KK}} $ and diameter $r_\mathrm{D}$. This representation is valid for $\xi_\mathrm{e} < r_\mathrm{D}$, which automatically implies $r_\mathrm{D} < \textbf{l}_\mathrm{\mathrm{KK}} $. Each quasi-monomer can be viewed as the rodlike array of $l_\mathrm{KK} / r_\mathrm{D}$ non-penetrating Debye blobs. Electrostatic and Debye blobs are shown in pink and black, respectively.}
\label{fig:C-QM}
\end{figure}

At higher salt concentrations, that is, lower Debye radii, PE exhibits rodlike statistics at the intermediate length, $\xi_\mathrm{e} < r < \textbf{l}_\mathrm{\mathrm{KK}}$. The PE chain can be considered as a coarse-grained chain of $N_\mathrm{KK} \simeq L_\mathrm{e} / \textbf{l}_\mathrm{\mathrm{KK}}$ quasi-monomers, each of length $\textbf{l}_\mathrm{\mathrm{KK}}$,
as schematically shown in Figure~\ref{fig:C-QM}.
Because these quasi-monomers repel each other in Coulomb interactions, they do not come closer than the Debye radius. Therefore, the effective diameter of the quasi-monomer is equal to $r_\mathrm{D}$, which is much lower than $\textbf{l}_\mathrm{\mathrm{KK}}$, and the second virial coefficient for them scales as $B_\mathrm{KK} \simeq \textbf{l}_\mathrm{\mathrm{KK}}^2 r_\mathrm{D}$, as for cylinders of length $\textbf{l}_\mathrm{\mathrm{KK}}$ and diameter $r_\mathrm{D}$. (In terms of Section~\ref{sec:EEV}, one may say that the quasi-monomer consists of $l_\mathrm{KK}/r_\mathrm{D}$ non-penetrating Debye blobs.)

To define whether or not excluded volume interactions between quasi-monomers lead to the chain swelling, one should estimate the respective excluded volume perturbation theory parameter:~\cite{GK-book}
\begin{equation}
    z_\mathrm{KK} \simeq \frac{B_\mathrm{KK}} {\textbf{l}_\mathrm{\mathrm{KK}}^3} N_\mathrm{KK}^{1/2} 
    \simeq 
    \frac{\textbf{l}_\mathrm{\mathrm{KK}}^2 r_\mathrm{D}}{\textbf{l}_\mathrm{\mathrm{KK}}^3} \left( \frac{L_\mathrm{e}}{\textbf{l}_\mathrm{\mathrm{KK}}} \right)^{1/2}
    \simeq \frac{ r_\mathrm{D} L_\mathrm{e}^{1/2} } { \textbf{l}_\mathrm{\mathrm{KK}}^{3/2} } 
    \simeq \frac{l_\mathrm{0} \xi_\mathrm{e} }{r_\mathrm{D}^2} N^{1/2} 
\label{eq:z_KK}
\end{equation}
PE chain starts swelling and exhibits global swollen-coil statistics at $z_\mathrm{KK} \simeq 1$, and the respective crossover can be written as $r_\mathrm{D}^{saPE1/saPE2} \simeq (\xi_\mathrm{e} / l_\mathrm{0})^{1/2} N^{1/4}$, or equivalently
\begin{equation}
     \left( u f^2 \right)_\mathrm{saPE1/saPE2} \simeq 
     \left( \frac{r_\mathrm{D}} {l_\mathrm{0}} \right)^{-6} N^{3/2}   
\end{equation}
In the scaling diagram constructed in Figure~\ref{fig:2}, this crossover is shown in orange. It takes place at lower $r_\mathrm{D}$ values compared to the sfPE/saPE1 crossover given by eq.~\ref{eq:cross-sfPE/saPE1}. 

Therefore, as salt is added and the Debye radius goes down, flexible PE first ceases to be globally rodlike (crossover sfPE/saPE1) and only later it starts to swell due to excluded volume interactions (crossover saPE1/saPE2). The regime saPE1 confined between these two boundaries corresponds to the globally ideal-coil PE statistics. The chain can be viewed ideal random walk of the $\textbf{l}_\mathrm{\mathrm{KK}} $ quasi-monomers with the end-to-end distance~\cite{KK-1982, RGJ-2024}
\begin{equation}
    R_\mathrm{saPE1} \simeq \textbf{l}_\mathrm{\mathrm{KK}}  N_\mathrm{KK}^{1/2} 
    \simeq l_\mathrm{0} \frac{r_\mathrm{D}} {\xi_\mathrm{e} } N^{1/2} 
    \simeq l_\mathrm{0} u^{-1/6} f^{2/3} c_s^{-1/2} N^{1/2}
\label{eq:R-saPE1}
\end{equation}

\subsubsection{Salt-Added PE with Swollen-Coil Global Statistics 
and Local Electrostatic Stiffening (Regime saPE2)}

Beyond the saPE1/saPE2 crossover, at lower $r_\mathrm{D}$ values, the chain of KK quasi-monomers starts to swell owing to $z_\mathrm{KK} \gg 1$. The swollen coil of quasi-monomers has the dimensions of
\begin{equation}
    R_\mathrm{saPE2} \simeq \textbf{l}_\mathrm{\mathrm{KK}} N_\mathrm{KK}^{1/2} z_\mathrm{KK}^{1/5}
    \simeq \textbf{l}_\mathrm{\mathrm{KK}} \left( \frac{B_\mathrm{KK}} {\textbf{l}_\mathrm{\mathrm{KK}}^3} \right)^{1/5} N_\mathrm{KK}^{3/5}
    \simeq \frac{l_\mathrm{0}^{6/5} r_\mathrm{D}^{3/5}}{ \xi_\mathrm{e}^{4/5}} N^{3/5} 
    \simeq l_\mathrm{0} u^{-1/30} f^{8/15} c_s^{-3/10} N^{3/5} 
\label{eq:saPE2}
\end{equation}
The saPE1/saPE2 crossover from ideal- to swollen-coil global statistics, given by $z_\mathrm{KK} \simeq 1$, can also be interpreted as the equality between the chain size $R$ and the size $\xi_\mathrm{T}^\mathrm{KK}$ of the thermal blob of KK quasi-monomers with respect to their excluded volume interactions: 
\begin{equation}
    \xi_\mathrm{T}^\mathrm{KK} 
    \simeq \textbf{l}_\mathrm{\mathrm{KK}}  \left( \frac{B_\mathrm{KK}} {\textbf{l}_\mathrm{\mathrm{KK}}^3} \right)^{-1} 
    \simeq \frac{r_\mathrm{D}^3} {\xi_\mathrm{e}^2} 
    \simeq l_\mathrm{0} u^{-5/6} f^{4/3} c_s^{-3/2}
\label{eq:xi-T-KK}
\end{equation}
Each KK thermal blob comprises $\left( B_\mathrm{KK} / \textbf{l}_\mathrm{\mathrm{KK}}^3 \right)^{-2}$ KK quasimonomers, which corresponds to
\begin{equation}
    g_\mathrm{T}^\mathrm{KK} 
    \simeq g_\mathrm{e} \frac{\textbf{l}_\mathrm{\mathrm{KK}} } {\xi_\mathrm{e}} \left( \frac{B_\mathrm{KK}} {\textbf{l}_\mathrm{\mathrm{KK}}^3} \right)^{-2}    
    \simeq \frac{r_\mathrm{D}^4}{ l_\mathrm{0}^2 \xi_\mathrm{e}^2}
    \simeq u^{-4/3} f^{4/3} c_s^{-2}
\label{eq:g-T-KK}
\end{equation}
bare statistical segments, and $N \simeq g_\mathrm{T}^\mathrm{KK}$ at the crossover.
The internal mean-squared distance of the chain in regime saPE2 is provided in the Supporting Information, Figure S3.

Regime saPE2 terminates at the crossover 
\begin{equation}
    \left( u f^2 \right)_\mathrm{saPE2/SC} \simeq \left( \frac{r_\mathrm{D}} {l_0} \right)^{-1} 
\label{eq:saPE2/SC}
\end{equation}
given by eqs.~\ref{triple-cross-KK}-\ref{triple-cross-KK-2}, which is shown in blue in Figure~\ref{fig:2}. Here, several changes (``aspects'') happen simultaneously:
\begin{enumerate}[label=(\alph*)]
\item The EPL equal to $\textbf{l}_\mathrm{\mathrm{KK}} $ no longer exceeds the size of the coarse-grained unit equal to the electrostatic blob $\xi_\mathrm{e}$, which limits the region of KK calculations.
\item Electrostatic stiffening for flexible chains becomes negligible, as shown in Figure~\ref{fig:1} where the respective blue solid line delineates the domains of bare stiffness and KK el-st stiffening. 
\item The cylindrical quasi-monomer of $\textbf{l}_\mathrm{\mathrm{KK}}  \times r_\mathrm{D}$ dimensions becomes spherical at this crossover, $\textbf{l}_\mathrm{\mathrm{KK}}  \simeq \xi_\mathrm{e} \simeq r_\mathrm{D}$.
Hence, Coulomb interactions between quasi-monomers, treated as an effective excluded volume, transition from anisotropic and orientation-dependent to isotropic.
\item At the crossover, Coulomb interactions between quasi-monomers become weak because the charge of each KK quasi-monomer is about the charge of the electrostatic blob, $q_\mathrm{e} \simeq e f g_\mathrm{e}$. Debye blobs change from non-penetrating to penetrating.
\end{enumerate}

\subsubsection{Swollen Coil with Electrostatic Excluded Volume 
and No Electrostatic Stiffening (Regime SC)}

Changes (a)-(d) enable considering screened Coulomb repulsions at high salt concentrations within the concept of electrostatic excluded volume discussed in Section~\ref{sec:EEV}. Point (d) enables linearization of the Boltzmann factor in calculations of the effective second virial coefficient due to the penetrating nature of Debye blobs, cf. eq.~\ref{eq:virial-DD}.  Using the result of eq.~\ref{eq:EEV-deriv} for $B_\mathrm{eev} \simeq l_\mathrm{0} u f^2 r_\mathrm{D}^2$, one can find the perturbation theory parameter for electrostatic excluded volume interactions:
\begin{equation}
    z_\mathrm{eev} \simeq \frac{B_\mathrm{eev}} {l_\mathrm{0}^3} N^{1/2} 
    \simeq \frac{ l_\mathrm{0} r_\mathrm{D}^2} {\xi_\mathrm{e}^{3}}  N^{1/2}   
\label{eq:z-eev}
\end{equation}
At the saPE2/SC crossover where $r_\mathrm{D} \simeq \xi_\mathrm{e}$, it exceeds unity by far, $z_\mathrm{eev} \simeq (N / g_\mathrm{e})^{1/2}$, provided that chains are not too short, $N \gg g_\mathrm{e}$.  

Therefore, the regime neighboring to saPE2 is indeed the swollen coil regime SC, in which the chain size is equal to
\begin{equation}
    R_\mathrm{SC} \simeq l_\mathrm{0} N^{1/2} z_\mathrm{eev}^{1/5} 
    \simeq l_\mathrm{0} \left( \frac{B_\mathrm{eev}} {l_0^3 } \right)^{1/5} N^{3/5} 
    \simeq \frac{ l_\mathrm{0}^{6/5} r_\mathrm{D}^{2/5}} { \xi_\mathrm{e}^{3/5} } N^{3/5}
    \simeq l_\mathrm{0} c_s^{-1/5} f^{2/5} N^{3/5}
\label{eq:R-SC}
\end{equation}

\subsubsection{Ideal Coil (Regime IC)}

The effect of the electrostatic excluded volume becomes negligible at $z_\mathrm{eev} \simeq 1$, that is, when the strength of Coulomb interactions $u f^2$ is lower than
\begin{equation}
    \left( u f^2 \right)_\mathrm{SC/IC} \simeq \left( \frac{r_\mathrm{D}}{l_\mathrm{0}} \right)^{-2}  N^{-1/2} 
\label{eq:SC/IC}
\end{equation}
This crossover is shown in Figure~\ref{fig:2} in gray and remains also valid for semiflexible PEs, as will be shown in the next subsection.
One can also identify SC/IC crossover with the equality between the chain size $R$ and the size of the thermal blob of bare monomers due to electrostatic excluded volume $\xi_\mathrm{T}^\mathrm{eev}$ given by
\begin{equation}
    \xi_\mathrm{T}^\mathrm{eev} \simeq l_\mathrm{0} \left( \frac{B_\mathrm{eev}} {l_\mathrm{0}} \right)^{-1}
    \simeq \frac{\xi_\mathrm{e}^3} {r_\mathrm{D}^2}
    \simeq l_\mathrm{0} f^{-2} c_s
\label{eq:xi-T-eev}
\end{equation}
At the crossover, the chain length $N$ is approximately equal to the number of bare monomers $g_\mathrm{T}^\mathrm{eev}$ in the thermal blob due to electrostatic excluded volume
\begin{equation}
    g_\mathrm{T}^\mathrm{eev} \simeq \left( \frac{B_\mathrm{eev}}{l_\mathrm{0}^3} \right)^{-2} 
    \simeq \frac{\xi_\mathrm{e}^6} {r_\mathrm{D}^4 l_\mathrm{0}^2}
    \simeq f^{-4} c_s^2
\label{eq:g-T-eev}
\end{equation}

The ideal coil regime with the chain size of
\begin{equation}
    R_\mathrm{IC} \simeq l_\mathrm{0} N^{1/2}
\end{equation}
is implemented at very low Debye radii, that is, at very high salt concentrations.

\begin{center}
\begin{table*}[h]
\caption{Scaling regimes and power laws for the PE chain size. The chain size $R$ is given in three different forms (representations). In form 2, the dimensionless salt concentration $c_s$ is the product of actual salt concentration $C_\mathrm{s}$ and cubed bare persistence length, $c_\mathrm{s} = C_\mathrm{s} l_\mathrm{0}^3$. }
\label{table:1}
\centering
\renewcommand{\arraystretch}{2.0}
\begin{tabular}{|c|c|c|c|c|}
\hline
Regime
&  Conformation type
&  Chain size $R$ (form 1)    
&  Chain size $R$ (form 2)   
&  Chain size $R$ (form 3) 
\\ \hline
IC      
&  ideal coil
&  $ l_\mathrm{0} N^{1/2} $ 
&  $ l_\mathrm{0} N^{1/2} $
&  $ l_\mathrm{0}^{1/2} L^{1/2} $
\\ \hline
SC
&  swollen coil
&  $ l_\mathrm{0}^{6/5} r_\mathrm{D}^{2/5} \xi_\mathrm{e}^{-3/5} N^{3/5} $
&  $ l_\mathrm{0} f^{2/5} c_s^{-1/5} N^{3/5}$
&  $ l_\mathrm{0}^{1/5} l_\mathrm{B}^{1/5} A^{-2/5} r_\mathrm{D}^{2/5} L^{3/5}  $
\\ \hline
saPE2
&  globally swollen coil
&  $ l_\mathrm{0}^{6/5} r_\mathrm{D}^{3/5} \xi_\mathrm{e}^{-4/5} N^{3/5} $
&  $ l_\mathrm{0} u^{-1/30} f^{8/15} c_s^{-3/10} N^{3/5} $
& $ l_\mathrm{0}^{1/15} l_\mathrm{B}^{4/15} A^{-8/15} r_\mathrm{D}^{3/5} L^{3/5} $
\\ \hline
saPE1
&  globally ideal coil
&  $ l_\mathrm{0} r_\mathrm{D} \xi_\mathrm{e}^{-1} N^{1/2} $
&  $ l_\mathrm{0} u^{-1/6} f^{2/3} c_s^{-1/2} N^{1/2} $
&  $ l_\mathrm{0}^{-1/6} l_\mathrm{B}^{1/3} A^{-2/3} r_\mathrm{D} L^{1/2} $
\\ \hline
sfPE
&  rodlike stretch
&  $ l_\mathrm{0}^2  \xi_\mathrm{e}^{-1} N $
&  $ l_\mathrm{0}  u^{1/3} f^{2/3} N $
&  $ l_\mathrm{0}^{1/3} l_\mathrm{B}^{1/3} A^{-2/3} L $
\\ \hline
OH
&  globally swollen coil
&  $ l_\mathrm{0}^{4/5} r_\mathrm{D}^{1/5} N^{3/5} $
&  $ l_\mathrm{0} u^{-1/10} c_s^{-1/10} N^{3/5} $  
&  $ l_\mathrm{0}^{1/5} r_\mathrm{D}^{1/5} L^{3/5} $
\\ \hline
$\overline{\text{saPE2}}$
&  globally swollen coil
&  $ l_\mathrm{0}^{3/5} r_\mathrm{D}^{3/5}  ( \xi_\mathrm{e}^{|} )^{-1/5} N^{3/5} $
&  $ l_\mathrm{0} u^{-1/10} f^{2/5} c_s^{-3/10} N^{3/5} $
&  $ l_\mathrm{B}^{1/5} A^{-2/5} r_\mathrm{D}^{3/5} L^{3/5} $
\\ \hline
$\overline{\text{saPE1}}$
&  globally ideal coil
&  $ l_\mathrm{0}^{1/2} r_\mathrm{D}  ( \xi_\mathrm{e}^{|} )^{-1/2} N^{1/2} $
&  $ l_\mathrm{0} f c_s^{-1/2} N^{1/2} $
&  $ l_\mathrm{B}^{1/2} A^{-1} r_\mathrm{D} L^{1/2} $
\\ \hline
$\overline{\text{sfPE}}$
&  rodlike stretch
&  $ l_\mathrm{0} N $
&  $ l_\mathrm{0} N $
&  $ L $
\\ \hline
\end{tabular}
\end{table*}
\end{center}

\subsection{Semiflexible Polyelectrolytes, $u f^2 \gg 1$}

In this section, we consider conformations of semiflexible PEs. Most of the conformational regimes --- $\overline{\text{sfPE}}$, $\overline{\text{saPE1}}$, and $\overline{\text{saPE2}}$ --- are analogous to those for flexible chains, so they are denoted the same way, with the extra bar added to emphasize that the chains are intrinsically stiff. The difference arises from two different crossovers present in Figure~\ref{fig:1} for semiflexible chains. This leads to the new conformational regime OH (described first by Odijk and Houwaart~\cite{odijk-1978}), in which electrostatic stiffening is already absent, but Coulombic interactions between quasi-monomers remain strong and anisotropic.

\subsubsection{Salt-Free PE with Rod-like Global Statistics (Regime $\overline{\text{sfPE}}$) } 

At very low salt concentrations, semiflexible PEs exhibit rodlike statistics at all length scales, locally due to the bare chain stiffness and globally due to long-range Coulomb repulsions. The chain size simply coincides with the contour length
\begin{equation}
    R_\mathrm{\overline{sfPE}} \simeq l_\mathrm{0} N
\end{equation}

This picture holds as long as the EPL exceeds the chain contour length. The crossover to the salt-added regimes is given by $\textbf{l}_\mathrm{\mathrm{OSF}} \simeq l_\mathrm{0} N$ and can be written as
\begin{equation}
    \left( u f^2 \right)_\mathrm{\overline{sfPE} / \overline{saPE1}} \simeq 
    \left( \frac{r_\mathrm{D}} {l_0} \right)^{-2} N
\end{equation}
It is shown with the red line in the top part of Figure~\ref{fig:2}.

\subsubsection{Salt-Added PE with Ideal Coil Global Statistics 
and Local Electrostatic Stiffening (Regime $\overline{\text{saPE1}}$)} 

When the bare chain stiffness is high, $u f^2 \gg 1$, and salt concentration is low, the electrostatic stiffening is described by the OSF theory predicting $\textbf{l}_\mathrm{\mathrm{OSF}} \simeq r_\mathrm{D}^2 / \xi_\mathrm{e}^{|}$. The following analysis is analogous to that for flexible chains. The equivalent chain representation is $N_\mathrm{OSF} \simeq l_\mathrm{0} N / \textbf{l}_\mathrm{\mathrm{OSF}}$ quasi-monomers, each of length $\textbf{l}_\mathrm{\mathrm{OSF}}$ and the effective diameter $r_\mathrm{D}$. (The quasi-monomer can be considered as the linear array of $\textbf{l}_\mathrm{\mathrm{OSF}} / r_\mathrm{D}$ Debye blobs, which are non-penetrating due to $r_\mathrm{D} > \xi_\mathrm{e}^{|}$ in this regime.) The excluded volume of quasi-monomers is $B_\mathrm{OSF} \simeq \textbf{l}_\mathrm{\mathrm{OSF}}^2 r_\mathrm{D}$, and the excluded volume interaction parameter is~\cite{odijk-1978}
\begin{equation}
    z_\mathrm{OSF} \simeq \frac{B_\mathrm{OSF}} { \textbf{l}_\mathrm{\mathrm{OSF}}^3 } N_\mathrm{OSF}^{1/2}
    \simeq \frac{ r_\mathrm{D} \left( l_\mathrm{0} N \right)^{1/2}} { \textbf{l}_\mathrm{\mathrm{OSF}}^{3/2} }
    \simeq \frac{ l_\mathrm{0}^{1/2} ( \xi_\mathrm{e}^{|} )^{3/2} } {r_\mathrm{D}^2} N^{1/2}
\label{eq:z-OSF}
\end{equation}
in similarity to the flexible PEs case, cf. eq.~\ref{eq:z_KK}.

By extending the analogy to flexible chains, it can be demonstrated that 
$z_\mathrm{OSF} \ll 1$ near the $\overline{\text{sfPE}}$/$\overline{\text{saPE1}}$ crossover. Therefore, PE conformations in regime $\overline{\text{saPE1}}$ can be viewed as ideal random walk of $N_\mathrm{OSF}$ steps, each of $\textbf{l}_\mathrm{\mathrm{OSF}}$ length, leading to the following chains size:
\begin{equation}
    R_\mathrm{\overline{saPE1}} \simeq \textbf{l}_\mathrm{\mathrm{OSF}} N_\mathrm{OSF}^{1/2} 
    \simeq \left( l_\mathrm{0} \textbf{l}_\mathrm{\mathrm{OSF}} N \right)^{1/2} 
    \simeq \frac{ l_\mathrm{0}^{1/2} r_\mathrm{D} } { (\xi_\mathrm{e}^{|})^{1/2} } N^{1/2} 
    \simeq l_\mathrm{0} f c_s^{-1/2} N^{1/2} 
\label{eq:R-saPE1-bar}
\end{equation}

\subsubsection{Salt-Added PE with Swollen Coil Global Statistics 
and Local Electrostatic Stiffening (Regime $\overline{\text{saPE2}}$)} 

Excluded volume interactions between the cylindrical OSF quasi-monomers set in at $z_\mathrm{OSF} \simeq 1$, that is, at
\begin{equation}
    \left( u f^2 \right)_\mathrm{\overline{saPE1} / \overline{saPE2}}
    \simeq \left( \frac{r_\mathrm{D}} {l_0} \right)^{-4/3} N^{1/3}
\end{equation}
This crossover is shown in orange in the top part of Figure~\ref{fig:2}. 
At the crossover, the PE chain comprises about one thermal OSF blob, $R \simeq \xi_\mathrm{T}^\mathrm{OSF}$ and $N \simeq g_\mathrm{T}^\mathrm{OSF}$. The latter corresponds to the chain size, above which the chain of OSF quasi-monomers swells due to excluded volume repulsions between them. In similarity to eqs.~\ref{eq:xi-T-KK}-\ref{eq:g-T-KK}, the OSF thermal blob size and the number of bare monomers in it are given by
\begin{equation}
    \xi_\mathrm{T}^\mathrm{OSF} 
    \simeq \textbf{l}_\mathrm{\mathrm{OSF}} \left( \frac{B_\mathrm{OSF}} {\textbf{l}_\mathrm{\mathrm{OSF}}^3} \right)^{-1} 
    \simeq \frac{r_\mathrm{D}^3} {( \xi_\mathrm{e}^{|} )^2} 
    \simeq l_\mathrm{0} u^{1/2} f^{4} c_s^{-3/2}
\label{eq:xi-T-OSF}
\end{equation}
\begin{equation}
    g_\mathrm{T}^\mathrm{OSF} 
    \simeq \frac{\textbf{l}_\mathrm{\mathrm{OSF}}} { l_\mathrm{0} } \left( \frac{B_\mathrm{KK}} {\textbf{l}_\mathrm{\mathrm{KK}}^3} \right)^{-2}   
    \simeq \frac{r_\mathrm{D}^4}{ l_\mathrm{0}^2 ( \xi_\mathrm{e}^{|} )^2}
    \simeq u f^{6} c_s^{-2}
\label{eq:g-T-OSF}
\end{equation}

In regime $\overline{\text{saPE2}}$, the chain is the swollen coil comprising $N_\mathrm{OSF}$ quasi-monomers, each of length $\textbf{l}_\mathrm{\mathrm{OSF}}$ and repelling each other with the second virial coefficient $B_\mathrm{OSF}$. In analogy to regime saPE2 and eq.~\ref{eq:saPE2}, its size is equal to
\begin{equation}
    R_\mathrm{\overline{saPE2}} \simeq \textbf{l}_\mathrm{\mathrm{OSF}} N_\mathrm{OSF}^{1/2} z_\mathrm{OSF}^{1/5} 
    \simeq \textbf{l}_\mathrm{\mathrm{OSF}} \left( \frac{B_\mathrm{OSF}} {\textbf{l}_\mathrm{\mathrm{OSF}}^3} \right)^{1/5} N_\mathrm{OSF}^{3/5}
    \simeq \frac{ l_0^{3/5} r_\mathrm{D}^{3/5} } {  ( \xi_\mathrm{e}^{|} )^{1/5} } N^{3/5}
    \simeq l_\mathrm{0} u^{-1/10} f^{2/5} c_s^{-3/10} N^{3/5}
\label{eq:R-saPE2}
\end{equation}
It deserves mentioning that the other choice of the quasi-monomer, with $l_\mathrm{0}$ length, would also lead to eqs.~\ref{eq:R-saPE1-bar} and~\ref{eq:R-saPE2}.~\cite{footnote-2}

The analogy between the flexible regimes sfPE, saPE1, and saPE2 and the their semiflexible analogues $\overline{\text{sfPE}}$, $\overline{\text{saPE1}}$, and $\overline{\text{saPE2}}$ is apparent. The respective scalings versus the chain length, equal to $1$, $3/5$, and $1/2$, are identical. The red crossovers between the salt-free and salt-added regimes 1 are given by the same physical requirement of equality between the EPL and the chain end-to-end distance. The orange crossovers between salt-added regimes 1 and 2 are also fully analogous and correspond to $z \simeq 1$ for quasi-monomers, the only difference being the quasi-monomer length choice, $\textbf{l}_\mathrm{KK} $ versus $\textbf{l}_\mathrm{OSF}$. Finally, the flexible/semiflexible crossovers between the respective regimes can be found from matching the chains' sizes (or setting $\textbf{l}_\mathrm{OSF} \simeq \textbf{l}_\mathrm{KK}$) and appears universal: $u f^2 \simeq 1$, or equivalently, $\xi_\mathrm{e} \simeq \xi_\mathrm{e}^{|} \simeq l_\mathrm{0}$, in agreement with eq.~\ref{eq:stiff/flex}. This additionally clarifies the convenience of identifying the rodlike electrostatic blob size $\xi_\mathrm{e}^{|}$ for semiflexible chains as the counterpart of the conventional Gaussian $\xi_\mathrm{e}$.

Regime $\overline{\text{saPE2}}$ terminates together with the OSF electrostatic stiffening dominance. The green crossover in Figure~\ref{fig:2}, given by $l_\mathrm{0} \simeq \textbf{l}_\mathrm{\mathrm{OSF}}$, or equivalently
\begin{equation}
    \left( u f^2 \right)_\mathrm{\overline{saPE2}/OH} \simeq \left( \frac{r_\mathrm{D}}{l_\mathrm{0}} \right)^{-2}, 
\label{eq:saPE2bar/OH}
\end{equation}
demarcates it from the domain of bare stiffness dominance, cf. Figure~\ref{fig:1}. This green crossover also signifies the termination of the complete analogy between flexible and semiflexible chains. 

It is instructive to compare it to the saPE2/SC crossover for flexible chains in the context of (a)-(d) changes happening in the latter case, see eq.~\ref{eq:saPE2/SC} and the discussion after it. For semiflexible chains, aspect (a) is irrelevant. Aspect (b) certainly does take place, as it defines the crossover $\overline{\text{saPE2}}$/OH from el-st to bare stiffness. However, neither (c) nor (d) aspects are relevant at this crossover: quasi-monomers remain cylinders with the length much higher than the diameter, $l_\mathrm{0} \gg r_\mathrm{D}$, and Coulomb interactions between them still by far exceed the thermal energy. These changes will only take place later, at the blue crossover given by $\textbf{l}_\mathrm{\mathrm{OSF}} \simeq r_\mathrm{D} \simeq \xi_\mathrm{e}^{|}$ and demarcating regime OH from SC and IC regimes. In other words, saPE2/SC crossover found for flexible chains \textit{splits into two} distinct crossovers, $\overline{\text{saPE2}}$/OH and OH/SC, for semiflexible chains. This opens a window of regime OH, which is unique for semiflexible PEs and is absent for their flexible counterparts.

\subsubsection{Salt-Added PE with Swollen Coil Global Statistics and No Electrostatic Stiffening (Regime OH)~\cite{odijk-1978}}

Below the crossover given by eq.~\ref{eq:saPE2bar/OH}, the Debye radius is sufficiently low to make electrostatic stiffening negligible. However, Coulomb interactions cannot yet be treated within the na\"{i}ve electrostatic excluded volume concept! This is discussed in Section~\ref{sec:EEV}. Indeed, near the $\overline{\text{saPE2}}$/OH crossover, Debye radius remains much higher than the rodlike electrostatic blob, $r_\mathrm{D} > \xi_\mathrm{e}^{|}$, suggesting strong, by far exceeding the thermal energy of $k_\mathrm{B} T$, Coulomb repulsions between the Debye blobs. That is to say, Debye blobs remain non-penetrating.

In the OH regime, the PE chain should be viewed as $N$ Kuhn segments, each of the bare persistence length $l_\mathrm{0}$ and the diameter $r_\mathrm{D}$. Note that the condition $A \ll r_\mathrm{D}$ is also assumed to be fulfilled in the OH regime, ensuring that the Debye blobs of adjacent charges overlap, thereby forming a continuous cylindrical shell around the chain. The resulting second virial coefficient scales as $B_\mathrm{OH} \simeq l_\mathrm{0}^2 r_\mathrm{D}$, in agreement with eq.~\ref{eq:B_OH} of Section~\ref{sec:EEV}, and the perturbation parameter is
\begin{equation}
    z_\mathrm{OH} \simeq \frac{l_\mathrm{0}^2 r_\mathrm{D}} {l_\mathrm{0}^3} N^{1/2} 
    \simeq \frac{r_\mathrm{D}} {l_\mathrm{0}} N^{1/2} 
\label{eq:zOH}
\end{equation}
The resulting scaling law for the chain size
\begin{equation}
    R_\mathrm{OH} \simeq l_\mathrm{0} N^{1/2} z_\mathrm{OH}^{1/5}
    \simeq l_\mathrm{0} \left( \frac{B_\mathrm{OH}}{l_\mathrm{0}^3} \right)^{1/5} N^{3/5} 
    \simeq l_0^{4/5} r_\mathrm{D}^{1/5} N^{3/5}
    \simeq l_\mathrm{0} u^{-1/10} c_s^{-1/10} N^{3/5}
\label{eq:R-OH}
\end{equation}
differs from those in $\overline{\text{saPE2}}$ and SC regimes. 

The idea to set the effective chain diameter equal to $r_\mathrm{D}$ belongs to Odijk and Houwaart,~\cite{odijk-1978} who arrived at the same result for $z_\mathrm{OH}$. Interestingly, the power law for $R_\mathrm{OH}$ has earlier been found by Dobrynin within the perturbation calculations in one of his latest papers on EPL,~\cite{DC-2009}
which resolved discrepancies with his earlier work.~\cite{dobrynin-2005}

The OH regime terminates and crossover to the SC regime takes place when repulsions between the rodlike chain fragments are so weak that they no longer provide non-penetration between Kuhn segments of $l_\mathrm{0} \times r_\mathrm{D}$ dimensions, or equivalently, between the Debye blobs that become penetrating. This happens when the energy of Coulomb repulsions between them is sufficiently weak to enable calculations of the electrostatic excluded volume $B_\mathrm{eev}$, see Section~\ref{sec:EEV}. For flexible chains, Coulomb interactions cease to be strong (non-linearizable) at $r_\mathrm{D} \simeq \xi_\mathrm{e}$. For stiff chains, the analogous condition reads 
\begin{equation}
    \textbf{l}_\mathrm{\mathrm{OSF}} \simeq r_\mathrm{D} \simeq \xi_\mathrm{e}^{|} 
\label{eq:OH/SC}
\end{equation}
and is shown with a dashed blue line in Figure~\ref{fig:2}. 

Formally, the first equality in eq.~\ref{eq:OH/SC} may seem unnecessary because there is no electrostatic stiffening in regime OH, $\textbf{l}_\mathrm{\mathrm{OSF}} \ll l_\mathrm{0}$, which makes $\textbf{l}_\mathrm{\mathrm{OSF}}$ seemingly irrelevant length. However, we intentionally write down this crossover in the triple form to emphasize the analogy to its counterpart for flexible chains, which is given by eq.~\ref{triple-cross-KK-2} and shown with solid blue lines in Figures~\ref{fig:1} and~\ref{fig:2}. 

Recall that eq.~\ref{eq:OH/SC} for semiflexible chains is \textit{not a complete} analog of eq.~\ref{triple-cross-KK-2} for flexible chains because the former is associated with aspects (c) and (d) only, while the latter also denotes termination of the electrostatic stiffening and therefore pertains to all four aspects, (a)-(d). For semiflexible chains, the transition to the bare stiffness domain, referred to as aspect (b), happens earlier, at the green crossover given by $\textbf{l}_\mathrm{\mathrm{OSF}} \simeq l_\mathrm{0}$, while aspect (a), namely theoretical coarse-graining to el-st blobs, appears fully irrelevant.

\subsubsection{Swollen Coil with Electrostatic Excluded Volume and No Electrostatic Stiffening (Regime SC)}

In the SC regime, semiflexible chains behave the same way as flexible chains. By applying the earlier performed analysis, one arrives at eq.~\ref{eq:R-SC}, which can also be written as
\begin{equation}
    R_\mathrm{SC} 
    \simeq \frac{ l_\mathrm{0}^{4/5} r_\mathrm{D}^{2/5}} { (\xi_\mathrm{e}^{|})^{1/5} } N^{3/5}
    \simeq l_\mathrm{0} f^{2/5} c_s^{-1/5} N^{3/5}
\end{equation}
The SC regime is common for flexible and semiflexible chains, and crossing $uf^2 \simeq 1$ boundary does not affect the scaling law for the chain dimensions.

To recapitulate, although both OH and SC regimes are referred to as the regimes of globally swollen-soil conformations with no electrostatic stiffening, they differ in the strength and anisotropy of Coulomb interactions between the Kuhn segments: strong anisotropic for OH and weak isotropic for SC. This manifests in non-penetrating and penetrating types of Debye blobs in these regimes, respectively.

\subsubsection{Ideal Coil (Regime IC)}

SC regime terminates when the electrostatic excluded volume parameter is on the order of unity, $z_\mathrm{eev} \simeq 1$, see eq.~\ref{eq:z-eev}. 
The resulting SC/IC crossover given by eq.~\ref{eq:SC/IC} is also common for flexible and semiflexible chains; it is shown with a gray line in Figure~\ref{fig:2}. At very low Debye radii, the PE conformations are Gaussian coils, $R_\mathrm{IC} \simeq l_\mathrm{0} N^{1/2}$.

Direct transition from  OH to IC regime is also possible. OH regime terminates when excluded volume interactions in it become weak, $z_\mathrm{OH} \simeq 1$, see eq.~\ref{eq:zOH}. The respective OH/IC crossover can be written as
\begin{equation}
    \left( \frac{ r_\mathrm{D} } { l_\mathrm{0} } \right)_\mathrm{IC/OH} \simeq N^{-1/2}
\end{equation}
and is shown with the vertical gray line in Figure~\ref{fig:2}. 
Equivalently, in the spirit of eqs.~\ref{eq:xi-T-eev}-\ref{eq:g-T-eev}, for flexible chains, one may introduce the effective thermal blob for OH regime with
\begin{equation}
    \xi_\mathrm{T}^\mathrm{OH} \simeq l_\mathrm{0}  \left( \frac{B_\mathrm{OH}} {l_\mathrm{0}^3} \right)^{-1}
    \simeq \frac{l_\mathrm{0}^2} {r_\mathrm{D}}
    \simeq l_\mathrm{0} u^{1/2} c_s^{1/2}
\label{eq:xi-T-OH}
\end{equation}
\begin{equation}
    g_\mathrm{T}^\mathrm{OH} \simeq  \left( \frac{B_\mathrm{OH}} {l_\mathrm{0}^3} \right)^{-2}
    \simeq \frac{l_\mathrm{0}^2} {r_\mathrm{D}^2}
    \simeq u c_s
\label{eq:g-T-OH}
\end{equation}
and argue that the chain comprises on the order of one blob at the OH/IC crossover.

One can check that the three crossover lines --- IC/OH, OH/$\overline{\text{saPE2}}$, and $\overline{\text{saPE2}}$/$\overline{\text{saPE1}}$ --- meet at the same point with the $r_\mathrm{D} / l_\mathrm{0} \simeq N^{-1/2}$ and $u f^2 \simeq N$ coordinates. This is where regimes OH and $\overline{\text{saPE2}}$ with the global swollen coil statistics vanish. Above this point, increasing the reduced Debye radius leads to the direct transition from quasi-neutral IC regime to $\overline{\text{saPE1}}$ regime with also global ideal-coil statistics but local el-st stiffening given by OSF result. Hence, the crossover between these two regimes is set by the equality between the bare and electrostatic persistence lengths, $l_\mathrm{0} \simeq \textbf{l}_\mathrm{\mathrm{OSF}}$, and coincides with the OH/$\overline{\text{saPE2}}$ crossover given by eq.~\ref{eq:saPE2bar/OH}; it is shown with the green line in the scaling diagram of Figure~\ref{fig:2}.

All discussed scaling regimes and crossovers between them are summarized in Figure~\ref{fig:2}, and Table~\ref{table:1} provides the respective scaling laws for the chain size. It is also convenient to represent the power laws in various forms, as functions of the bare stiffness $l_\mathrm{0}$ and other different characteristic lengths/parameters of the problem: (i) in form 1, as the function of the electrostatic blob sizes, $\xi_\mathrm{e}$ and $\xi_\mathrm{e}^{|}$, the Debye radius $r_\mathrm{D}$; (ii) in form 2, as the function of the reduced salt concentration $c_s$ and electrostatic parameters, $f$ and $u$; (iii) in form 3, as the function of the Bjerrum length $l_\mathrm{B}$, the distance between the charges $A$, and the chain contour length $L$.

\section{Discussion and Conclusions}
\label{sec:conclusions}


In this paper, the problem of the electrostatic persistence length (EPL) is reconsidered to construct a complete diagram of conformational regimes for a single-chain polyelectrolyte (PE) in a salt-added $\Theta$-solvent solution. The resulting diagram shown in Figure~\ref{fig:2} unifies semiflexible and flexible PEs, highlighting both similarities and differences between them. In both cases, the EPL scales quadratically with the Debye radius, $\textbf{l}_\mathrm{e} \sim r_\mathrm{D}^{2}$, as first proposed by the Odijk–Skolnick–Fixman (OSF) and Khokhlov–Khachaturian (KK) theories for semiflexible and flexible PEs, respectively. These scaling laws are fully supported by the results of coarse-grained Monte Carlo and molecular dynamics simulations performed in the accompanying paper~\cite{EPL-part2}, which tested the end-to-end distance, internal mean-squared distances, and the decay of orientational correlations of the PE chain.

To properly identify different conformational regimes for semiflexible chains, a new characteristic length scale --- termed the rodlike electrostatic blob $\xi_\mathrm{e}^{|}$ and defined by eq.~\ref{eq:xi-rod-def} --- has been introduced. Using this length allows the OSF and KK results to be represented in a unified form: $\textbf{l}_\mathrm{\mathrm{OSF}} \simeq r_\mathrm{D}^2 / \xi_\mathrm{e}^{|}$ and $\textbf{l}_\mathrm{\mathrm{KK}}  \simeq r_\mathrm{D}^2 / \xi_\mathrm{e}$. This formulation also enables a generalization to PEs with arbitrary local conformational statistics characterized by the exponent $\nu$, yielding the electrostatic persistence length in the form $\textbf{l}_\mathrm{e}^{(\nu)} \simeq r_\mathrm{D}^2 / \xi_\mathrm{e}^{(\nu)}$.

The key difference in conformational regimes between semiflexible and flexible PEs arises at high salt concentrations, when electrostatic stiffening vanishes. In flexible chains, the onset of the bare stiffness domain occurs when Coulombic interactions become weak, linearizable, and effectively isotropic. In semiflexible chains, however, these changes upon salt addition do not occur simultaneously: the dominance of bare stiffness emerges before strong screening of Coulomb repulsions sets in. This opens a window for the special Odijk–Houwaart (OH) regime, which is absent in flexible PEs.

The scaling picture of the PE across all identified regimes is essential for adequate analysis of their conformations, which provides a consistent definition of the EPL. Namely, the EPL should reflect only \textit{local electrostatic stiffening}, rather than the effective excluded-volume interactions that produce a globally swollen coil statistics with the exponent $\nu \approx 3/5$. This distinction is crucial for the {\it consistent extraction} of the EPL from simulations and, especially, from experiments.~\cite{review-dilute-PE} In other words, approximating the PE chain as an ideal coil with Kuhn segment $\textbf{l}_\mathrm{e}$ is only justified in saPE1 and $\overline{\text{saPE1}}$ regimes, and not the others. In addition, PEs should be sufficiently long to clearly exhibit the scaling slopes of the respective regimes, as discussed in detail in the accompanying paper.~\cite{EPL-part2} These aspects are crucial for experimental testing of the EPL scalings and comparing them to theory. 

By confirming, both theoretically and through simulations, the quadratic dependence of the EPL for flexible chains, our results have important implications for understanding other PE systems. In classical theories of semidilute solutions~\cite{DCR-1995, DR-2005} and salt-added brushes of flexible PEs,~\cite{ZR-2012} a linear dependence of the EPL on the Debye radius, $\textbf{l}_\mathrm{e} \sim r_\mathrm{D}$, has been implicitly employed, and these theories have been particularly successful in rationalizing experimental data. 
A potential explanation for this apparent contradiction lies in the asymptotic nature of the KK scaling. Simulations in the accompanying paper~\cite{EPL-part2} demonstrate that the KK slope of 2 in the $\textbf{l}_\mathrm{\mathrm{KK}}  \sim r_\mathrm{D}^2$ can be most clearly distinguished from the linear Barrat-Joanny (BJ) scaling, $\textbf{l}_\mathrm{\mathrm{BJ}} \sim r_\mathrm{D}$, in the saPE1 regime only (see Supporting Information). However, this regime is sufficiently wide and the respective dependence becomes clearly visible only for sufficiently long chains, $N \approx 10^3$ and higher, provided that all monomers are charged. For shorter chains with $N \approx 10^2$, the apparent slope for the effective EPL is closer to 1, since the relevant scaling regime is too narrow for the higher slope of $2$ to fully develop. Given that in real experimental studies of PE solutions the length of PE strands within the correlation length (mesh size) rarely exceeds $N \approx 10^2$, it is plausible that the phenomenological linear dependence has worked so well. Experimental studies on solutions of long PEs at low (yet still semidilute) concentrations may shed further light on this issue and stimulate the development of more precise (non-scaling) theories of the EPL, which should, however, reproduce the scaling results of the present work in the respective limiting cases.

One should also keep in mind that, in experimental systems, effects such as counterions condensation/adsorption (including the role of counterion type),~\cite{brilliantov-1998, deshkovski-2001, andelman-2003, dobrynin-2006, dobrynin-2007, rumyantsev-2014, DR-2006-neck} polymer/solvent dielectric mismatch,~\cite{KK-1994, KK-1996, muthu-2004, rumyantsev-2013} charge and backbone solvation,~\cite{lin-2022} finite backbone thickness,~\cite{trizac-2016} as well as potentially poor solvent quality for the polymer backbone,~\cite{khokhlov-1980, DRO-1996, dobrynin-2007} may all also play an important role. They may render the conformational behavior less universal than predicted by the scaling theory developed herein and supported by simulations with implicit salt ions and solvent reported in the accompanying article.~\cite{EPL-part2} 
For instance, poor-solvent PEs can be described by adapting the approach of ref.~\citenum{RGJ-2024} for charge-imbalanced polyampholytes, i.e., by replacing the blob size set by Coulomb attractions with the thermal blob governed by hydrophobic interactions. Another aspect concerns a quantitatively accurate description of the chain’s finite extensibility. Within the developed scaling analysis, the chain was assumed to be locally Gaussian in flexible-chain regimes or fully stretched rodlike in semiflexible-chain regimes. These are limiting behaviors that are independent of the chain flexibility mechanism --- wormlike or freely jointed --- which makes the derived scalings robust. Constructing an accurate interpolation between these two limiting scalings to describe chain behavior at strong but yet incomplete stretching would require a careful choice of the chain flexibility model (Hamiltonian) for semiflexible chains.~\cite{DR-2010} Incorporating these effects into the present theory is left for future work.

We believe that the present work provides a comprehensive understanding of the EPL in both flexible and semiflexible chains and lays the foundation for refining the scaling picture of semidilute PE solutions.


\section*{Appendix: Logarithmic Corrections to Scaling Laws}

\setcounter{equation}{0}
\setcounter{subsection}{0}
\renewcommand{\theequation}{A.\arabic{equation}} 
\renewcommand{\thesubsection}{A.\arabic{subsection}} 

In the main text, for the sake of methodological clarity and simplicity, we neglected logarithmic corrections to the power laws for the chain dimensions. These corrections arise from two independent effects: (a) stronger-than-linear stretching of flexible PEs due to long-range Coulomb repulsions, and (b) electrostatic excluded volume at strong Coulomb repulsions. The first correction is relevant in the flexible regimes sfPE, saPE1, and saPE2. The second correction applies to the regimes saPE2, $\overline{\text{saPE2}}$, and OH, which exhibit globally swollen-coil statistics with strong and anisotropic excluded volume. Notably, saPE2 is the only regime where both effects contribute simultaneously. No logarithmic corrections arise in the regimes IC, SC, $\overline{\text{saPE1}}$, and $\overline{\text{sfPE}}$. The resulting logarithmic factors are summarized in Table~\ref{table:2}, and a brief derivation is provided below.

\begin{center}
\begin{table*}[h]
\caption{Logarithmic multiplicative corrections to the scaling laws of Table~\ref{table:1}. Dash (---) indicates that no logarithmic correction is required for the scaling power law. }
\label{table:2}
\centering
\renewcommand{\arraystretch}{2.0}
\begin{tabular}{|c|c|c|}
\hline
Regime
&  Conformation type
&  Correction
\\ \hline
IC    
& ideal coil
&  ---
\\ \hline
SC
& swollen coil
&  ---
\\ \hline
saPE2
& globally swollen coil
& $  \left[ \ln \left( r_\mathrm{D} / \xi_\mathrm{e} \right) \right]^{4/15} $
\\ \hline
saPE1
& globally ideal coil
& $ \left[ \ln \left( r_\mathrm{D} / \xi_\mathrm{e} \right) \right]^{-1/6}  $
\\ \hline
sfPE
&  rodlike stretch
&  $ \left[ \ln \left( N / g_\mathrm{e} \right) \right]^{1/3} $
\\ \hline
OH
&  globally swollen coil
&  $ \left[ \ln \left( r_\mathrm{D} / \xi_\mathrm{e}^{|} \right) \right]^{1/5} $
\\ \hline
$\overline{\text{saPE2}}$
&  globally swollen coil
&  $ \left[ \ln \left( r_\mathrm{D} / \xi_\mathrm{e}^{|} \right) \right]^{1/5} $
\\ \hline
$\overline{\text{saPE1}}$
& globally ideal coil
&  ---
\\ \hline
$\overline{\text{sfPE}}$
& rodlike stretch
&  ---
\\ \hline
\end{tabular}
\end{table*}
\end{center}

\subsection*{A. Stronger-Than-Linear Stretching of Flexible Polyelectrolytes}

\textit{Regime sfPE.}
For salt-free flexible PEs, this effect has been known since de Gennes~\cite{degennes-1976}, who derived it from balancing the Coulomb energy of the chain, considered as a cylinder of unknown length $R$ and known thickness $\xi_\mathrm{e}$ given by eq.~\ref{elst-blob}, and its elastic energy:
\begin{equation}
    \frac {F_\mathrm{tot}(R)} {k_\mathrm{B} T} \simeq \frac{R^2} {l_\mathrm{0}^2 N} 
    + \frac{l_\mathrm{B} f^2 N^2}{R} \ln \left( \frac{R}{\xi_\mathrm{e}} \right) 
\label{eq:A1}
\end{equation}
Solving $\partial F_\mathrm{tot} / \partial R = 0$ leads to stronger-than-linear stretching, as confirmed in simulations:~\cite{degennes-1976, DR-2005}
\begin{equation}
    R_\mathrm{sfPE}^\mathrm{Log} \simeq \xi_\mathrm{e} \frac{N}{g_\mathrm{e}} 
    \left[ \ln \left( \frac{N}{g_\mathrm{e}} \right) \right]^{1/3}
    \simeq l_\mathrm{0} \left( u f^2 \right)^{-1/3} N 
    \left[ \ln \left( u^{2/3} f^{4/3} N \right)  \right]^{1/3}
\label{eq:R_sfPE_ln}
\end{equation}
Here, the superscript ``Log'' denotes a more exact, corrected limiting law.
The ratio under the logarithm can also be written as $N / g_\mathrm{e} \simeq R / \xi_\mathrm{e}$, where the chain size $R$ represents the range of Coulomb repulsions between the most distant parts of the chain. Note that here, the result is expressed in terms of the usual, non-corrected parameters of the el-st blob: $\xi_\mathrm{e}$ given by eqs.~\ref{elst-blob} and $g_\mathrm{e} \simeq \xi_\mathrm{e}^2 / l_\mathrm{0}^2$.

\textit{Regime saPE1.}
In the presence of salt, the effective range of Coulomb repulsions controlling chain \textit{stretching} (but not its stiffening) is equal to the Debye radius.~\cite{dobrynin-2005} The above reasoning and eq.~\ref{eq:R_sfPE_ln} can now be applied to the chain fragment of length $r_\mathrm{D}$ (Debye blob): 
\begin{equation}
    r_\mathrm{D} \simeq \xi_\mathrm{e} \frac{g_\mathrm{D}^\mathrm{Log}} {g_\mathrm{e}}
    \left[ \ln \left( \frac{r_\mathrm{D}}{\xi_\mathrm{e}} \right) \right]^{1/3}
\end{equation}
This enables to find the corrected number $g_\mathrm{D}^\mathrm{Log}$ of monomers in it
\begin{equation}
    g_\mathrm{D}^\mathrm{Log} \simeq g_\mathrm{e} \frac{r_\mathrm{D}} {\xi_\mathrm{e}} \left[ \ln \left( \frac{r_\mathrm{D}}{\xi_\mathrm{e}} \right) \right]^{-1/3} \simeq g_\mathrm{D} \eta^{-1/3} < g_\mathrm{D} 
\end{equation}
which is lower than the non-corrected $g_\mathrm{D} \simeq g_\mathrm{e} r_\mathrm{D} / \xi_\mathrm{e}$. Here, the correcting multiplier
\begin{equation}
    \eta \equiv \ln \left( \frac {r_\mathrm{D}}  {\xi_\mathrm{e}} \right) > 1
\label{eq:eta}
\end{equation}
has been introduced. The physical reason for the diminution of the number of monomers in the Debye blob is the stronger-than-linear PE stretching. This results in a decrease in the effective linear charge density of the chain of el-st blobs, $\widetilde{f}^\mathrm{Log} = \widetilde{f} \eta^{-1/3}$ with the non-corrected $\widetilde{f}$ given by eq.~\ref{eq:renorm}, and the respective decrease in the KK EPL, $\textbf{l}_\mathrm{\mathrm{KK}}^\mathrm{Log} = \textbf{l}_\mathrm{\mathrm{KK}}  \eta^{-2/3}$. 

In contrast, the contour length of the chain of the el-st blob increases, 
$L_\mathrm{e}^\mathrm{Log} = r_\mathrm{D} N / g_\mathrm{D}^\mathrm{Log} = L_\mathrm{e} \eta^{1/3}$. Using eq.~\ref{eq:R-saPE1}, one can see that the combination of these two effects leads to
\begin{equation}
    R_\mathrm{saPE1}^\mathrm{Log} 
    \simeq l_\mathrm{0} \frac{r_\mathrm{D}}{\xi_\mathrm{e}} N^{1/2}  
    \left[ \ln \left( \frac{r_\mathrm{D}}{\xi_\mathrm{e}} \right) \right]^{-1/6} 
    \simeq R_\mathrm{saPE1} \eta^{-1/6}
\label{eq:R-saPE1-Log}
\end{equation}
The reduction of the chain size in regime saPE1 is due to the shortening of KK EPL ($\eta^{-2/3}$ factor), dominating the elongation of the chain of el-st blobs ($\eta^{1/3}$ factor). 

However, correcting the two scales, $\textbf{l}_\mathrm{\mathrm{KK}} $ and $L_\mathrm{e}$, both due to effect (a), is not sufficient to find the entire logarithmic correction for regime saPE2. This is because effect (b) also manifests in this regime. For this reason, we first consider the role of the latter in isolation.

\subsection*{B. Correction to Electrostatic Excluded Volume at Strong Repulsions}

Let us first consider the general problem of two $Z$-valent ions interacting via a screened Coulomb potential
\begin{equation}
     \frac {u_\mathrm{12} (r)} {k_\mathrm{B} T} = l_\mathrm{B} Z^2 \frac{ e^{-r/r_\mathrm{D}} }{r}
\end{equation}
Calculating the effective second virial coefficient of their repulsions defined by eq.~\ref{eq:virial} can be performed by dividing the integral into two parts: (a) the core part with $r \leq D$ where interactions are strong, $u_\mathrm{12} / k_\mathrm{B} T \gg 1$, and the exponent in the Mayer function is asymptotically equal to zero; (b) the tail part where interactions become weak, $u_\mathrm{12} / k_\mathrm{B} T \ll 1$ and the exponent can be linearized. The core part radius can be found from $ u_\mathrm{12} / k_\mathrm{B} T = 1$ and equals
\begin{equation}
    D \simeq r_\mathrm{D} \ln \left( Z^2 \frac{l_\mathrm{B}} {r_\mathrm{D}} \right)
\label{eq:D-vs-Z}
\end{equation}
It exceeds $r_\mathrm{D}$ provided that the charge is sufficiently high, $Z^2 l_\mathrm{B} / r_\mathrm{D} > 1$ --- an inequality defining strong, non-linearizable Coulomb repulsions. Note that in the analogous integration of the main text, eqs.~\ref{eq:virial-DD} and \ref{eq:B_eev_S}, $D$ was set to $r_\mathrm{D}$, i.e., the logarithmic factor was neglected. Retaining it here leads to a logarithmic correction to the second virial coefficient 
\begin{equation}
    B_\mathrm{ZZ}^\mathrm{Log}
    \simeq \int_\mathrm{0}^{D} d^3 r + l_\mathrm{B} Z^2 \int_\mathrm{D}^{\infty} e^{-r/r_\mathrm{D}} r dr
    \simeq D^3 + r_\mathrm{D} D^2 \simeq D^3 
    \simeq r_\mathrm{D}^3 \left[ \ln \left( Z^2 \frac{l_\mathrm{B}} {r_\mathrm{D}} \right) \right]^3  
\label{eq:B-ZZ}
\end{equation}
This refines the result of the main text, where $B_\mathrm{ZZ} \simeq r_\mathrm{D}^3$, and results in similar corrections to the chain size in regimes with strong electrostatic excluded volume: OH, $\overline{\text{saPE2}}$, and saPE2.

\textit{Regime OH.} The above results imply that, due to Coulomb repulsions, two OH quasi-monomers strongly ($\gg k_\mathrm{B} T$) repel each other and have an effective diameter (thickness) equal to $D (Z)$. Setting $Z$ to the charge of the Debye blob, $Z \simeq r_\mathrm{D} / A  \simeq  f r_\mathrm{D} / l_\mathrm{0}$, one can find
\begin{equation}
    D_\mathrm{OH} \simeq r_\mathrm{D} \ln \left( \frac{u f^2 r_\mathrm{D}} {l_\mathrm{0}} \right) 
    \simeq r_\mathrm{D} \ln \left( \frac{r_\mathrm{D}}{ \xi_\mathrm{e}^{|} } \right) 
    \simeq r_\mathrm{D} \eta^{|} 
\end{equation}
This result is in agreement with more precise calculations specifying numerical coefficients.~\cite{fixman-1978-part1, odijk-1986, rods-numerics-2023}.
In analogy to eq.~\ref{eq:eta}, we have introduced the correction factor for semiflexible chains:
\begin{equation}
    \eta^{|} = \ln \left( \frac{r_\mathrm{D}}{ \xi_\mathrm{e}^{|} } \right) > 1
\end{equation}
The resulting second virial coefficient for OH quasimonomers of $l_\mathrm{0} \times r_\mathrm{D} \eta^{|}$ size is equal to $B_\mathrm{OH} \simeq l_\mathrm{0}^2 r_\mathrm{D} \eta^{|}$. Combined with eq.~\ref{eq:R-OH}, this leads to the chain size of
\begin{equation}
    R_\mathrm{OH}^\mathrm{Log} \simeq l_\mathrm{0}^{4/5} r_\mathrm{D}^{1/5} N^{3/5} 
    \left[ \ln \left( \frac{r_\mathrm{D}}{ \xi_\mathrm{e}^{|} } \right) \right]^{1/5} \simeq R_\mathrm{OH} \left[ \eta^{|} \right]^{1/5}
\end{equation}

\textit{Regime $\overline{\text{saPE2}}$.} Exactly the same reasoning can be applied to regime $\overline{\text{saPE2}}$, the only difference being the length of the quasi-monomer equal to $\textbf{l}_\mathrm{\mathrm{OSF}}$ rather than $l_\mathrm{0}$. Since no corrections to $\textbf{l}_\mathrm{\mathrm{OSF}}$ are expected, one can readily find $B_\mathrm{OSF}^\mathrm{Log} \simeq \textbf{l}_\mathrm{\mathrm{OSF}}^2 D_\mathrm{OH} \simeq B_\mathrm{OSF} \eta^{|}$ and, according to eq.~\ref{eq:R-saPE2}, the resulting chain size 
\begin{equation}
    R_\mathrm{\overline{saPE2}}^\mathrm{Log} 
    \simeq  \frac{ l_0^{3/5} r_\mathrm{D}^{3/5} } {  ( \xi_\mathrm{e}^{|} )^{1/5} } N^{3/5}
    \left[ \ln \left( \frac{r_\mathrm{D}}{ \xi_\mathrm{e}^{|} } \right) \right]^{1/5}
    \simeq R_\mathrm{\overline{saPE2}} \left[ \eta^{|} \right]^{1/5}  
\end{equation}
It is not surprising that the logarithmic corrections in regimes OH and $\overline{\text{saPE2}}$ coincide --- they simply account for the effective thickness of the chain increasing from $r_\mathrm{D}$ to $r_\mathrm{D} \eta^{|}$.

\textit{Regime saPE2.} In this regime, both effects (a) and (b) act simultaneously. It was earlier discussed in the context of regime saPE1 that the chain overstretching leads to the corrected values of $\textbf{l}_\mathrm{\mathrm{KK}}^\mathrm{Log} = \textbf{l}_\mathrm{\mathrm{KK}}  \eta^{-2/3}$ and $L_\mathrm{e}^\mathrm{Log} = L_\mathrm{e} \eta^{1/3}$. To account for effect (b), one should remember that now the Debye blob consists of Gaussian el-st blobs, and the respective effective valency of the charge in eq.~\ref{eq:D-vs-Z} is equal to $Z \simeq f g_\mathrm{e} r_\mathrm{D} / \xi_\mathrm{e} \simeq f r_\mathrm{D} \xi_\mathrm{e} / l_\mathrm{0}^2$. Here, we neglect logarithmic corrections to $Z$ because we do not seek $\ln ( \ln)$ corrections to the chain size. As a result, the effective diameter of the KK quasi-monomer in regime saPE2 equals 
\begin{equation}
    D_\mathrm{saPE2} \simeq r_\mathrm{D} \ln \left( \frac{r_\mathrm{D}} {\xi_\mathrm{e}} \right) \simeq r_\mathrm{D} \eta
\end{equation}
The effective electrostatic excluded volume of KK quasimonomers, $B_\mathrm{KK}^\mathrm{Log} \simeq \left( \textbf{l}_\mathrm{\mathrm{KK}}^\mathrm{Log} \right)^2  D_\mathrm{saPE2} \simeq B_\mathrm{KK} \eta^{-1/3}$, decreases, while the respective perturbation theory parameter increases as $z_\mathrm{KK}^\mathrm{Log} \simeq z_\mathrm{KK} \eta^{13/6}$. Using the result of eq.~\ref{eq:R-saPE1-Log} one can find that
\begin{equation}
    R_\mathrm{saPE2}^\mathrm{Log} \simeq \left( z_\mathrm{KK}^\mathrm{Log} \right)^{1/5} R_\mathrm{saPE1}^\mathrm{Log} 
    \simeq \frac{l_\mathrm{0}^{6/5} r_\mathrm{D}^{3/5}}{ \xi_\mathrm{e}^{4/5}} N^{3/5} 
    \left[ \ln \left( \frac{r_\mathrm{D}}{\xi_\mathrm{e}} \right) \right]^{4/15} 
    \simeq R_\mathrm{saPE2} \eta^{4/15}  
\end{equation}

The summary of all the logarithmic corrections is provided in Table~\ref{table:2}. The logarithmic shifts in the positions of some of the crossovers are straightforward and can be found by matching the corrected results for the chain size. It is appropriate to mention here that other regimes may also have some corrections, but additive and hence asymptotically small rather than logarithmic multiplicative. For instance, in regime $\overline{\text{sfPE}}$, the Coulomb term will also have $\ln$ factor, cf. eq.~\ref{eq:A1}, but this will not lead to the $\ln$ correction to the chain size because of the finite (already non-Gaussian) extensibility of the chain~\cite{GK-book, marko-siggia-1995, RC-book, D-book, DcR-2010} at $uf^2 \gg 1$.

Finally, when testing logarithmic corrections, one should consider the exact value of $\nu = 0.588$ in the leading power-law dependence, rather than $\nu \approx 3/5$ used throughout this work.


\section*{Nomenclature}
Abbreviations
\begin{description}
[style=nextline, leftmargin=2.5cm, labelwidth=2.3cm, align=left, itemsep=-5pt]
   \item[EPL] electrostatic persistence length
   \item[eev] electrostatic excluded volume
   \item[el-st] electrostatic
   \item[KK] Khokhlov-Khachaturian
   \item[OH]  Odijk-Houwaart
   \item[OSF] Odijk-Skolnik-Fixman
\end{description}

Notation in Formulas
\begin{description}
[style=nextline, leftmargin=2.5cm, labelwidth=2.3cm, align=left, itemsep=-5pt]
  \item[$A$] chain contour length between the adjacent charges
  \item[$\widetilde{A}$] effective distance between charged monomers on rodlike chain of Gaussian electrostatic blobs
  \item[$B_\mathrm{cc}$] second virial coeff. for charge-charge interaction
  \item[$B_\mathrm{DD}$] second virial coeff. of interaction of Debye blobs
  \item[$B_\mathrm{eev}$] effective second virial coeff. for monomer-monomer interaction (electrostatic excluded volume)
  \item[$\widetilde{B}_\mathrm{eev}^{S}$] effective second virial coeff. for monomer-monomer interactions at strong interactions
  \item[$B_\mathrm{KK}$] second virial coeff. of interaction of KK quasi-monomers of $\textbf{l}_\mathrm{\mathrm{KK}} \times r_\mathrm{D}$ size
  \item[$B_\mathrm{OH}$] second virial coeff. for quasimonomers of OH quasi-monomers of $l_\mathrm{0} \times r_\mathrm{D}$ size
  \item[$B_\mathrm{OSF}$] second virial coeff. of interaction of OSF quasi-monomers of $\textbf{l}_\mathrm{\mathrm{OSF}} \times r_\mathrm{D}$ size
  \item[$C_s$] salt concentration
  \item[$c_s \equiv C_s l_\mathrm{0}^3 $] reduced (dimensionless) salt concentration
  \item[$D$] effective quasi-monomer/chain diameter due to electrostatic excluded volume 
  \item[$d$] chain diameter
  \item[$\delta \theta$] bending angle
  \item[$e$] elementary charge
  \item[$\epsilon$] solvent dielectric constant
  \item[$\eta$, $\eta^{|}$] logarithmic corrections for flexible and semiflexible chains
  \item[$f$] fraction of bare Kuhn segments carrying $e$ charge 
  \item[$g_\mathrm{e}$] number of monomers in Gaussian electrostatic blob 
  \item[$g_\mathrm{e}^{(\nu)}$] number of monomers in generalized electrostatic blob with $\nu$-statistics 
  \item[$g_\mathrm{e}^{|}$] number of monomers in rodlike electrostatic blob 
  \item[$g_\mathrm{D}$] number of monomers in Debye blob
  \item[$g_\mathrm{T}^\mathrm{eev}$] number of monomers in thermal blob due to electrostatic excluded volume
  \item[$g_\mathrm{T}^\mathrm{eev}$] number of monomers in thermal blob in OH regime
  \item[$g_\mathrm{T}^\mathrm{KK}$] number of monomers in thermal blob for chain of KK quasi-monomers
  \item[$g_\mathrm{T}^\mathrm{OSF}$] number of monomers in thermal blob for chain of OSF quasi-monomers
  \item[$k_\mathrm{B} T$] thermal energy
  \item[$l_\mathrm{B}$] Bjerrum length
  \item[$L = l_\mathrm{0} N$] chain contour length 
  \item[$L_\mathrm{e}$] contour length of chain of Gaussian electrostatic blobs
  \item[$\textbf{l}$] total persistence length of chain
  \item[$l_\mathrm{0}$] bare Kuhn segment length (in the absence of charges)
  \item[$\textbf{l}_\mathrm{e}$] electrostatic persistence length
  \item[$\textbf{l}_\mathrm{e}^{(\nu)}$] generalized EPL for chain of generalized electrostatic blobs $\xi_\mathrm{e}^{(\nu)}$ 
  \item[$\textbf{l}_\mathrm{\mathrm{KK}}$] KK EPL for flexible chains
  \item[$\textbf{l}_\mathrm{\mathrm{OSF}}$] OSF EPL for semiflexible chains
  \item[$N$] number of bare Kuhn segments, each of $l_\mathrm{0}$ length, in chain
  \item[$N_\mathrm{KK}$] number of KK quasi-monomers in chain
  \item[$N_\mathrm{OSF}$] number of OSF quasi-monomers in chain
  \item[$\nu$]  scaling exponent; assumed $1/2$, $3/5$, $1$ for Gausssian, swollen-coil, and rodlike statistics
  \item[$q_\mathrm{D}$] charge of Debye blob
  \item[$q_\mathrm{e}$] charge of Gaussian electrostatic blob
  \item[$q_\mathrm{e}^{(\nu)}$] charge of generalized electrostatic blob 
  \item[$q_\mathrm{e}^{|}$] charge of rodlike electrostatic blob
  \item[$r$] pairwise distance
  \item[$r_\mathrm{D}$] Debye radius (length) of screening due to salt
  \item[$R_\mathrm{X}$] characteristic chais size in scaling regime X
  \item[$R_\mathrm{X}^\mathrm{Log}$] chain size in regime X with account for logarithmic correction
  \item[$u \equiv l_\mathrm{B} / l_\mathrm{0}$] reduced (dimensionless) Bjerrum length
  \item[$\textbf{u}$] unit tangent vector of chain
  \item[$u_\mathrm{12}(r)$] pairwise interaction potential
  \item[$(uf^2)_\mathrm{X/Y}$] crossover position between regimes X and Y
  \item[$W_\mathrm{D}$] energy of unscreened Coulomb repulsion between two Debye blobs
  \item[$\widetilde{f}$] effective fraction of charged monomers on rodlike chain of Gaussian el-st blobs 
  \item[$\xi_\mathrm{e}$] size of Gaussian electrostatic blob 
  \item[$\xi_\mathrm{e}^{(\nu)}$] size of generalized electrostatic blob size 
  with $\nu$-statistics in it
  \item[$\xi_\mathrm{e}^{|}$] size of rodlike electrostatic blob
  \item[$\xi_\mathrm{T}^\mathrm{eev}$] size of thermal blob due to electrostatic excluded volume
  \item[$\xi_\mathrm{T}^\mathrm{OH}$] size of thermal blob in OH regime
  \item[$\xi_\mathrm{T}^\mathrm{KK}$] size of thermal blob size for chain of KK quasi-monomers
  \item[$\xi_\mathrm{T}^\mathrm{OSF}$] size of thermal blob size for chain of OSF quasi-monomers
  \item[$Z$] valency of multivalent ion
  \item[$z_\mathrm{eev}$] excluded volume parameter for chain of monomers with el-st excluded volume
  \item[$z_\mathrm{KK}$] excluded volume parameter for chain of KK quasi-monomers
  \item[$z_\mathrm{OH}$] excluded volume parameter for chain of OH quasi-monomers
  \item[$z_\mathrm{OSF}$] excluded volume parameter for chain of OSF quasi-monomers
\end{description}


\section*{Acknowledgment}
A.M.R. and A.A.G. gratefully acknowledge Michael Rubinstein for detailed discussions. They also thank Andrey Dobrynin for useful conversations.

\end{document}